\documentclass[showpacs,preprintnumbers,amsmath,amssymb,aps,eqsecnum]{revtex4}
\usepackage[dvips]{graphics,color}
\usepackage{latexsym}
\usepackage{graphicx}
\def\be{\begin{equation}}
\def\ee{\end{equation}}
\def\bea{\begin{eqnarray}}
\def\eea{\end{eqnarray}}

\def\dd{{\rm d}}
\def\pd{\partial}
\def\arcsinh{{\,\text{arcsinh}}}

\begin{document}

\title{Primordial perturbations from slow-roll inflation on a brane}

\author{Kazuya Koyama$^1$, Andrew Mennim$^1$, V.~A.~Rubakov$^2$, David Wands$^1$
 and Takashi Hiramatsu$^3$ \\~}

\affiliation{$^1$Institute of Cosmology and Gravitation, University of
  Portsmouth, Portsmouth~PO1~2EG, United Kingdom\\
  $^2$Institute for Nuclear Research of the Russian Academy of Sciences,
  60$^{\,th}\!$ October Anniversary Prospect 7a, Moscow 117312, Russia \\
  $^3$Department of Physics, School of Science, University of Tokyo, 7-3-1 Hongo, Bunkyo, Tokyo 113-0033, Japan}

\date{\today}

\begin{abstract}
In this paper we quantise scalar perturbations in a
Randall-Sundrum-type model of inflation where the inflaton field is
confined to a single brane embedded in five-dimensional anti-de Sitter
space-time. In the high energy regime, small-scale inflaton fluctuations
are strongly coupled to metric perturbations in the bulk and
gravitational back-reaction has a dramatic effect on the behaviour of
inflaton perturbations on sub-horizon scales. 
This is in contrast to the standard four-dimensional result where 
gravitational back-reaction can be neglected on small scales. 
Nevertheless, this does not give rise to significant particle
production, and the correction to the power spectrum of the curvature
perturbations on super-horizon scales is shown to be suppressed by a
slow-roll parameter. We calculate the complete first order slow-roll
corrections to the spectrum of primordial curvature perturbations.

\end{abstract}


\maketitle

\section{Introduction}

Recent developments have suggested that our
four-dimensional Universe could lie on a brane embedded in a
higher-dimensional space-time~\cite{Review}. This new perspective
may dramatically change our picture of the early Universe. One of the
most studied scenarios is the model proposed by Randall and Sundrum
where a single brane is embedded in  five-dimensional anti-de Sitter
(adS) space-time~\cite{RS99}. In this model, the Friedmann equation is
modified at high energy densities~\cite{BDEL} and the bulk gravity is strongly
coupled to the dynamics on the brane.  Because of that 
inflation in the early Universe, and the perturbations generated during inflation, 
could be
significantly modified compared with conventional four-dimensional
models.

The simplest way to realize inflation in the brane world is to
consider inflation driven by the potential energy of a scalar
field, the inflaton, confined to the brane~\cite{MWBH99}.
The amplitude of the resulting scalar curvature perturbations is given by
\begin{equation}
 \langle {\cal R}_c^2 \rangle^{1/2}
 = \left(\frac{H}{\dot{\phi}} \right) \langle \delta \phi^2 \rangle^{1/2},
 \label{curvature0}
\end{equation}
where $H$ is the Hubble parameter, $\dot{\phi}$ is the time
derivative of the inflaton $\phi$, and $\delta \phi$ is the
inflaton fluctuation on a spatially flat
hypersurface. The quantum expectation value of the
inflaton fluctuations on super-horizon scales  in the de Sitter space-time
is 
\begin{equation}
\langle \delta \phi^2 \rangle = \left(\frac{H}{2 \pi} \right)^2.
\label{phide}
\end{equation}
It has been shown that the constancy of the curvature perturbation
${\cal R}_{c}$ on comoving hypersurfaces outside the horizon is
independent of the gravitational theory~\cite{WMLL} and thus it
 is also
valid in this brane-world model~\cite{LMSW}. The curvature
perturbation, Eq.~(\ref{curvature0}), can thus be directly related to
the observables like Cosmic Microwave Background (CMB) temperature
anisotropy. Although the formula for the curvature perturbation is
exactly the same as in the four-dimensional case, the relation between
the Hubble parameter $H$ and the scalar field potential $V$ is
modified at high energy densities due to the modification of the Friedmann
equation~\cite{MWBH99}. This changes the
prediction of the spectrum of scalar perturbations for a given
potential.  Based on Eq.~(\ref{curvature0}),  interesting progress
has been made on observational predictions of the brane-world
inflation models~\cite{obs,RL}.

A crucial assumption in deriving Eq.~(\ref{phide}), and hence
calculating the amplitude of the curvature perturbation
(\ref{curvature0}), is that back-reaction due to metric perturbations
in the bulk can be neglected.  In the extreme slow-roll
limit, where the coupling between inflaton fluctuations and metric
perturbations vanishes, this assumption must be valid, but it is not
obvious whether this assumption is still valid once slow-roll
effects are taken into account. Ref.~\cite{KLMW} initiated the
study of bulk metric perturbations generated by inflaton fluctuations
on a brane. It was shown that sub-horizon
inflaton fluctuations on
a brane excite an infinite ladder of Kaluza--Klein modes of the bulk
metric perturbations already at the first order in slow-roll
parameters. Subsequently Ref.~\cite{KMW} found that, due to this
infinite ladder of modes, a naive slow-roll expansion breaks down in the
high energy regime once one takes into account the back-reaction of the bulk
metric perturbations. This was confirmed in Ref.~\cite{HK} by direct
numerical simulations that solved the coupled equations for inflaton
perturbations and bulk metric perturbations without using any
slow-roll approximation.

Although these results indicate that the back-reaction of the bulk
metric perturbations give rise to corrections of order one to the
behaviour of inflaton fluctuations on sub-horizon scales even in slow-roll
inflation, this does not necessarily imply that the amplitude of the
inflaton perturbations receives corrections of order one on large scales. In
order to address the amplitude of the inflaton perturbations, and
hence the curvature perturbations (\ref{curvature0}) on large scales, we
need  to quantise properly 
the coupled brane inflaton fluctuations and
bulk metric perturbations.  The purpose of this paper is to perform 
quantisation of the coupled fluctuations and determine the corrections
to the inflaton fluctuations amplitude, Eq.~(\ref{phide}).  In this
way we study whether
the bulk metric perturbations give rise to a significant modification
of the amplitude of the primordial fluctuations, Eq.~(\ref{curvature0}),
generated during high-energy inflation on a brane.

This paper is organised as follows. In section II,
we begin by describing the model and present
the coupled equations for the inflaton fluctuations
on a brane and the metric perturbations in the bulk.
Slow-roll parameters are defined and our
approximation for dealing with the brane-bulk
coupling is explained. Based on this approximation,
late-time solutions are derived. In section III,
we study the behaviour of the bound states,
which determine the final amplitude of the
curvature perturbations. We perform numerical
simulations to evolve the late-time bound states
backwards in time and to find their early-time behaviour.
We also develop a semi-analytic method
to derive solutions for the bound states at early times.
The bound states are shown to consist of an infinite
ladder of the Kaluza--Klein modes. Using the semi-analytic
analysis, asymptotic solutions
for perturbations on small scales are derived.
Section IV is devoted to the quantisation of the bound
states. We present a method to determine the quantum amplitude
of fluctuations with the knowledge of asymptotic
solutions for the bound states on small scales.
Then the corrections to the formula (\ref{curvature0})
at the first order in the slow-roll parameters are presented and compared
with the standard slow-roll corrections in four-dimensional gravity.
We end up with our conclusions in Section~V.


\section{Description of the Model}
\subsection{Basic equations}
We take as a starting point the analyses 
of Refs.~\cite{KLMW,KMW,HK}.  The model under consideration is the single-brane 
Randall--Sundrum model~\cite{RS99} with a cosmological constant in the bulk and an 
inflaton field $\phi$ confined to the brane, whose potential $V(\phi)$ satsifies the 
slow-roll conditions.  The action for the model is
\begin{equation}
 S=\frac{1}{2\kappa_5^2} \int_{\cal M} d^5x \sqrt{-^{^{(5)}}\!g} \:\Big\{ R-2\Lambda \Big\}
 + \int_{\pd \cal M} d^4x \sqrt{-^{^{(4)}}\!g}
 \left\{ -\lambda + \frac{1}{\kappa_5^2}K \right\}
 - \int d^4 x \sqrt{-^{^{(4)}}\!g} \left\{ \frac{1}{2} (\pd\phi)^2 + V(\phi)\right\},
\end{equation}
where $K$ is the trace of the extrinsic curvature of the boundary ${\pd \cal M}$
and $\lambda$ is the tension. The adS curvature scale is defined by
$\Lambda = -6 \mu^2$ and the tension is tuned as 
$\lambda = 6 \mu /\kappa_5^2$.

The background line element has the form
\begin{equation}
 ds^2 =  - N^2(y,t)dt^2 + A^2(y,t)\delta_{ij}dx^i dx^j + dy^2.
 \label{eq:background_metric}
\end{equation}
Here the warp function $A(y,t)$ and the lapse function $N(y,t)$ are~\cite{BDEL}
%
\begin{align}
 A(y,t) &= a(t)\left\{\cosh (\mu y) -
 \left(1+\frac{\kappa_5^2\rho}{6\mu}\right)\sinh (\mu y) \right\},\\
 N(y,t) &= \cosh (\mu y) -
 \left\{1-\frac{\kappa_5^2}{6\mu}(2 \rho+3p)\right\}\sinh(\mu y),
 \label{eq:lapse_warp}
\end{align}
where
\begin{equation}
\rho = \frac{1}{2} \dot{\phi}^2 + V(\phi), \quad
p = \frac{1}{2} \dot{\phi}^2 - V(\phi).
\end{equation}
In these coordinates, the brane is located at $y=y_{\rm b}=0$.
The scale factor is determined by the
Friedmann equation and  the equation of motion for the scalar field,
%
\begin{eqnarray}
  H^2 &=& \frac{\kappa_4^2}{3}\rho\left(1+\frac{\rho}{2\lambda}\right),
 \label{eq:Friedmann_equation} \quad
 \ddot{\phi} + 3 H \dot{\phi} + V'(\phi)=0.
\end{eqnarray}
%
Allowing for the linearised metric perturbations gives the metric
\begin{equation}
  ds^2 = -N(y,t)^2(1+2 {\cal A}) dt^2+A(y,t)^2(1+2 {\cal R})\delta_{ij} dx^i dx^j\\
+(1+2 A_{yy})dy^2+N(y,t)A_{y}dydt.
  \label{eq:perturbed_metric}
 \end{equation}
%
It was shown in Refs~\cite{Mukohyama, Kodama} that the perturbed five-dimensional
Einstein equations are solved
in the adS background if the metric perturbations are derived
from a master variable $\Omega$:
%
\begin{align}
{\cal A} &=  -\frac{1}{6A}
    \left\{ \left( 2\Omega'' - \frac{N'}{N} \Omega' \right)
     + \frac{1}{N^2} \left( \ddot\Omega - \frac{\dot N}{N} \dot\Omega \right)
     - \mu^2 \Omega
    \right\}\, ,  \label{eq:metric_A}
    \\
 A_y &=
     \frac{1}{NA}\left( \dot\Omega'-\frac{N'}{N}\dot\Omega \right)\, ,
     \label{eq:metric_Ay} \\
 A_{yy} &= \frac{1}{6A} \left\{
     \left( \Omega''- 2\frac{N'}{N} \Omega' \right)
      + \frac{2}{N^2} \left( \ddot\Omega - \frac{\dot{N}}{N} \dot\Omega\right)
      + \mu^2 \Omega \right\} \,, \label{eq:metric_Ayy} \\
{\cal R} &=
    \frac{1}{6A}\left\{ \left( \Omega''+ \frac{N'}{N}\Omega' \right)
      +\frac{1}{N^2}\left(-\ddot\Omega + \frac{\dot N}{N} \dot\Omega \right)
      - 2\mu^2 \Omega\right\} \, . \label{eq:metric_R}
\end{align}
The remaining perturbed five-dimensional Einstein equation then gives
the wave equation in the bulk,
%
\begin{equation}
 - \left( \frac{1}{NA^3} \dot\Omega \right)^\cdot
 + \left( \frac{N}{A^3} \Omega' \right)^\prime
 + \left( \mu^2 - \frac{k^2}{A^2} \right)
 \frac{N}{A^3} \Omega = 0 \,,
 \label{eq:master_equation}
\end{equation}
for perturbations with comoving wavenumber $k$ on the brane.

On the brane, there is also the perturbation of the inflaton $ \delta \phi$.
It is useful to express the scalar field perturbations $\delta \phi$
on the brane in terms of the gauge invariant, Mukhanov--Sasaki 
variable~\cite{Mukhanov} 
%
\begin{equation}
  Q \equiv \delta \phi - \frac{\dot{\phi}}{H} {\cal R}_{\rm b},
  \label{eq:Mukhanov-Sasaki}
\end{equation}
where subscript b denotes  the quantity  evaluated at
the brane.
The equation of motion is~\cite{HK}
\begin{equation}
  \ddot{Q} + 3 H \dot{Q} + \frac{k^2}{a^2} Q
  + \left\{\frac{\ddot{H}}{H} -2 \frac{\dot{H}}{H} \frac{V'(\phi)}{\dot{\phi}}
  -2 \left(\frac{\dot{H}}{H}\right)^2 + V''(\phi) \right\}Q =J(\Omega),
  \label{eq:Mukhanov_Sasaki_equation}
\end{equation}
%
where
%
\begin{equation}
J(\Omega)=-\frac{\dot{\phi}}{H}
\left[
\left(- \frac{\dot{H}}{H} + \frac{\ddot{H}}{2 \dot{H}} \right)
\kappa_4^2 \delta q_{{\cal E}} + \frac{1}{3}
\left(1- \frac{\dot{H}}{2 {\cal H}^2} \right) \kappa_4^2 \delta \rho_{{\cal E}}
+ \frac{1}{3} \kappa_4^2 k^2 \delta \pi_{{\cal E}}
+ \frac{1}{3} \frac{\dot{H}}{{\cal H}^2} \frac{k^2}{a^2} {\cal R}_{\rm b}
\right],
  \label{eq:Mukhanov_Sasaki_source}
\end{equation}
${\cal H} = (A'/A)_{\rm b}$ and $\kappa_4^2 = \kappa_5^2 \mu$. 
Here
\begin{align}
  \kappa_4^2 \delta \rho_{\cal E} &= \left(
     \frac{k^4 \Omega}{3a^5}\right)_{\rm b},
     \label{eq:Weyl_1}\\
  \kappa_4^2 \delta q_{{\cal E}} &=-
     \frac{k^2}{3a^3}\left(H\Omega - \dot{\Omega} \right)_{\rm b},
     \label{eq:Weyl_2}\\
  \kappa_4^2 \delta \pi_{{\cal E}} &=
     \frac{1}{6a^3}\left\{3 \ddot{\Omega} - 3H\dot{\Omega}
    +\frac{k^2}{a^2}\Omega-3\left(\frac{N'}{N}-\frac{A'}{A}\right)
     \Omega' \right\}_{\rm b}
     \label{eq:Weyl_3}
\end{align}
come form the projected Weyl tensor~\cite{SMS, Deffayet}, while
the curvature perturbation, Eq.~(\ref{eq:metric_R}), evaluated at the brane
is given in terms of $\Omega$ as
%
\begin{equation}
{\cal R}_{\rm b} = \frac{1}{6a}
\left(3{\cal H} \Omega' -3H \dot{\Omega} - 3 \mu^2 \Omega
+ \frac{k^2}{a^2} \Omega \right)_{\rm b}.
  \label{eq:curvature_brane}
\end{equation}
The junction condition, which describes how the brane field provides
the boundary condition for the bulk metric perturbation is~\cite{HK}
\begin{equation}
 \kappa_5^2a\dot{\phi}^2\left(\frac{H}{\dot{\phi}}Q\right)^{\cdot}
 =-  \frac{k^2}{a^2}\left\{
    \frac{\kappa_5^2\dot{\phi}^2}{6}\left(\dot{\Omega}-H\Omega\right)
          + H\left(\Omega'-\frac{A'}{A}\Omega\right)
  \right\}_{\rm b}.
  \label{eq:junction}
\end{equation}
The master equation (\ref{eq:master_equation}), the
Mukhanov--Sasaki equation (\ref{eq:Mukhanov_Sasaki_equation}) with a
source (\ref{eq:Mukhanov_Sasaki_source}) and the junction condition
(\ref{eq:junction})
describe the coupled system of the inflaton and metric perturbations.

\subsection{Slow-roll inflation}
The slow-roll parameters are defined as
\begin{equation}
\epsilon_{_H} =-\frac{\dot{H}}{H^2}, \quad
\eta_{_H} = - \frac{\ddot{\phi}}{H \dot{\phi}} \; .
\end{equation}
In terms of the scalar potential, these parameters are given by~\cite{MWBH99}
\begin{eqnarray}
\epsilon_{_H} &=& \frac{1}{2 \kappa_4^2}
\left(\frac{V'}{V} \right)^2
\left[\frac{2 \lambda (2 \lambda + 2 V)}{(2 \lambda+V)^2}\right], \\
\eta_{_H} &=& \frac{1}{\kappa_4^2} \left(
\frac{V''}{V} \right) \left( \frac{2 \lambda}{2 \lambda +V} \right)
-\epsilon_{_H} \; .
\end{eqnarray}
Self-consistency of the slow-roll approximation requires
$\epsilon_{_H}, |\eta_{_H}|  \ll 1$. In the high energy regime,
$V \gg \lambda$ (equivalently $H/\mu \gg 1$),
the modification of the Friedmann
equation eases the condition for slow-roll inflation
for a given potential.

The amplitude of the scalar perturbations is determined
by the curvature perturbations on a
comoving hypersurface, which is conserved on large scales:
\begin{equation}
{\cal R}_c = -\frac{H}{\dot{\phi}} Q.
\end{equation}
At the zeroth order in the slow-roll parameters, the quantum expectation
value of squared $Q$ is $\langle Q^2 \rangle = (H/2 \pi)^2$.
Slow-roll dynamics of the inflaton field gives corrections to this
result. One contribution comes from the evolution of $Q$ on
super-horizon scales. As is seen from
Eq.~(\ref{eq:Mukhanov_Sasaki_equation}), deviations from the de Sitter
space-time give an effective mass to $Q$ and modify the behaviour of $Q$
outside the horizon.  In four-dimensional cosmology this correction is
known as the Stewart--Lyth correction. This Stewart--Lyth correction in the
present brane-world inflation model was calculated in Ref.~\cite{RL}
assuming that the coupling to the bulk metric perturbations could be
neglected.

In four-dimensional cosmology, the slow-roll corrections do not affect
the dynamics of $Q$ on sub-horizon scales,
as the effective mass
induced by the dynamics of the inflaton can be neglected in comparison with
the gradient term. It is then possible to quantise $Q$
as an effectively massless field in the de Sitter space-time.  However, in
the brane-world case there is no reason to believe that the coupling
to the bulk metric perturbations can be neglected on sub-horizon
scales. In fact in Refs.~\cite{KMW, HK} it is shown that the
coupling to the bulk metric perturbations can lead to $O(1)$ effect on
the behaviour of $Q$ on small scales at high energies ($H/\mu \gg 1$)
even though the coupling is suppressed by the slow-roll
parameter. However, it has been an open question how this coupling
affects the quantum amplitude of $Q$ on larger scales. The aim of this
paper is to investigate the effect  which the mixing between the bulk
metric perturbations and the brane inflaton fluctuations has on the
quantum amplitude of $Q$ after crossing out the horizon.  
We are only interested in corrections of
the first order in the slow-roll parameters. Thus, we can study
separately the effect of the brane-bulk coupling and other slow-roll
corrections. Almost to the end of this paper we  
neglect all the slow-roll corrections in the
equations of motion except for the coupling between $Q$ and $\Omega$,
as we study the effect of the brane-bulk coupling.

\subsection{de Sitter approximation}
As explained above, we assume zeroth-order slow-roll
for the background everywhere except for the terms inducing
the mixing between
the bulk and brane perturbations. This means that, in practice, we consider
the background as the de Sitter brane configuration~\cite{Kaloper},
\begin{equation}
\label{lineelement}
\dd s^2=N(z)^2\left[\dd z^2+\frac{1}{H^2\tau^2}\big(-\dd \tau^2+
\delta_{ij} dx^{i} dx^{j} \big)\right]\,,
\end{equation}
where
\begin{equation}
\label{Ndef}
N(z)=\frac{H}{\mu \sinh (Hz)}\,,
\end{equation}
and the brane is located at
\begin{equation}
z_{\rm b}=H^{-1}\arcsinh \left(\frac{H}{\mu} \right)\,.
\end{equation}
Here, $H$ is the Hubble parameter (constant in the de Sitter Universe)
and $\tau$ is the conformal time coordinate. Hereafter we redefine the 
time variable 
\begin{equation}
\tau \to - \tau \, ,
\end{equation}
so that evolution forward in $\tau$ is evolution backwards in time.

It is convenient to define a new variable $\chi$, related to the master variable as
follows,
\begin{equation}
\Omega=k^{-2} a^2 N^{3/2} \chi \,.
\end{equation}
The equation of motion for $\chi$  derived from Eq.~(\ref{eq:master_equation}) is
\begin{equation}
\label{eom:chi}
\chi_{,\tau \tau}+k^2\chi-\frac{2}{\tau^2}\chi+\frac{1}{H^2\tau^2}\big(-\chi''+U(z)\chi\big) = 0\,,
\end{equation}
where 
\begin{equation}
 U(z)=\frac{H^2}{4}\left( 9 -\frac{1}{\sinh^2(Hz)} \right) \,.
\end{equation}
Note that Eq.~(\ref{eom:chi}) is separable.
The other equations are simplified by defining a new brane variable
\begin{equation}
\label{def:u}
 u= aQ + \frac{\dot\phi}{6H} \chi|_{_{\rm b}} \,.
\end{equation}
It obeys the following equation of motion,
\begin{equation}
 \label{eom:u}
 u_{,\tau \tau}+k^2 u - \frac{2}{\tau^2} u
 = \frac{\dot\phi}{3H \tau } \left[\chi_{,\tau}
 -\frac{2}{\tau} \chi \right]_{_{\rm b}}
  \,,
\end{equation}
derived from Eq.~(\ref{eq:Mukhanov_Sasaki_equation}), and the junction condition
\begin{equation}
 \label{junction:u}
 \left[ \chi' + \left(\frac{1}{2}\frac{N'}{N}+ \frac{\kappa_5^2 \dot{\phi}^2}{3}  
 \right) \chi
 \right]_{_{\rm b}}
 = \kappa_5^2 \dot\phi H \left[\tau u_{,\tau} - u\right],
\end{equation}
derived from Eq.~(\ref{eq:junction}).
Here we have neglected all terms suppressed by the slow-roll
parameters except for the terms responsible for the coupling
between $u$ and $\chi$, i.e., the terms that induce the mixing between
the brane and bulk modes.
The latter terms at the first non-trivial order in slow-roll are $O(\beta)$, where
\begin{equation}
 \beta^2=\frac{\kappa_5^2 \dot{\phi}^2}{6H}.
\end{equation}
To the leading order in the slow-roll parameters, $\beta^2$ can be
written as
\begin{equation}
\beta^2= \frac{1}{3} \epsilon_{_H} 
\left( 1+\left(\frac{\mu}{H} \right)^2 \right)^{-1/2} \; .
\end{equation}
This is the parameter that controls the
strength of the coupling between the inflaton perturbation on the brane
and the gravitational
perturbations in the bulk.

The second order action for the variables $\Omega$ and  $Q$ 
has been derived in Ref.~\cite{YK}. In terms of $u$ and $\chi$, the action is
\begin{eqnarray}
 S&=&\int d^4x \left[\frac12 (u_{,\tau}^2-k^2u^2
 + \frac{2}{\tau^2} u^2)+\frac{\dot\phi}{3H}\left(\frac{u \chi_{,\tau}}{\tau}
 -\frac{2u\chi}{\tau^2} \right) \right]
 \nonumber\\
&+&
  \frac{1}{6\kappa_5^2} \int d^4x dz\left[\chi_{,\tau}^2-k^2\chi^2 + \frac{2}{\tau^2}
  \chi^2
 -\frac{{\chi'}^2 + U(z)\chi^2}{(H\tau)^2}\right] 
 +\frac{1}{6 \kappa_5^2} 
 \left[  \frac{1}{2}
 \left( \frac{N'}{N} \right)_{_{\rm b}} + \frac{\kappa_5^2 \dot{\phi}^2}{3}
 \right]
 \int d^4 x \frac{1}{(H \tau)^2}
 \chi^2\, .
\end{eqnarray}
From this action, the coupled equations (\ref{eom:chi}), (\ref{eom:u}) and (\ref{junction:u})
can be derived. It is useful to define the time-independent Wronskian of two solutions,
\begin{equation}
 W(f^*,f)=u^* u_{,\tau}-u u_{,\tau}^*-\frac{\dot\phi}{3H\tau}\big(u^*\chi|_{_{\rm b}}-u\chi^*|_{_{\rm b}}\big)
 +\frac{1}{3\kappa_5^2} \int^\infty_{z_{_{\rm b}}} dz\, \big(\chi^* \chi_{,\tau}
 - \chi \chi_{,\tau}^*\big) \,,
\label{wronskian}
\end{equation}
where $f=(u, \chi)$ is a state comprising a brane and a bulk mode.
The Wronskian will play an important role in  quantisation.

Without mixing, $u$ and $\chi$ would behave as canonical scalar fields in
four dimensions and five dimensions, respectively. The existence of the mixing terms in 
the field equations and in the Wronskian
makes the analysis complicated. We  simplify the equations further by defining yet
another new variable to describe the perturbations of the inflaton field on the brane,
\begin{equation}
 \label{def:xi}
 \xi =
 u_{,\tau} + \frac{1}{\tau} u - \frac{\dot{\phi}}{3H}
 \frac{1}{\tau}\chi|_{_{\rm b}}
 \,,
\end{equation}
which is related to the density perturbation on a comoving hypersurface~\cite{Kodama}.
The brane equation of motion (\ref{eom:u}) is now written as
\begin{equation}
 \label{eom:xi}
 \xi_{,\tau \tau}+k^2\xi=-\frac{k^2\dot\phi}{3 H \tau} \chi|_{_{\rm b}}\,,
\end{equation}
and the junction condition (\ref{junction:u}) becomes
\begin{equation}
 \label{junction:xi}
 \left[ \chi' +\frac{1}{2} \frac{N'}{N}  \chi \right]_{_{\rm b}}
 = \kappa_5^2 \dot{\phi} H \tau \xi \,.
\end{equation}
Upon substituting (\ref{def:xi}) into (\ref{wronskian}),
the Wronskian  becomes
\begin{equation}
W(f^*,f)=\frac{1}{k^2} (\xi^*\xi_{,\tau}-\xi \xi_{,\tau}^*)
 +\frac{1}{3\kappa_5^2} \int^\infty_{z_{_{\rm b}}} dz\, \big(\chi^* \chi_{,\tau}
 - \chi \chi_{,\tau}^*\big) \,,
\label{wr33}
\end{equation}
where $f=(\xi,\chi)$.

In this and next sections 
we are going to study the coupled system of equations (\ref{eom:chi}), 
(\ref{eom:u}) and (\ref{junction:u}), or, equivalently,  (\ref{eom:chi}),
(\ref{eom:xi}) and (\ref{junction:xi}).

\subsection{Late time solutions}
At late times, it is possible to find analytic
solutions to the coupled equations.
Solutions divide into  bound states and continuum
modes. There are two independent bound state solutions~\cite{KLMW}.
A growing bound state solution is 
\begin{equation}
 u_g=\frac{A_g}{\tau}+B\tau \,, \quad
 \chi_g= C_g \,\tau \sqrt{\sinh(Hz)} \, \log \left(\frac{\cosh(Hz)-1}{\cosh(Hz)+1}\right)\,,
\label{boundg}
\end{equation}
where $A_g$ is the overall amplitude, while $B$ and $C_g$ can be determined through $A_g$. 
In terms of $\xi$, the growing mode solution is given by
\begin{equation}
\xi_g = k^2 A_g,
\label{xig}
\end{equation}
which is clearly the growing mode solution of Eq.~(\ref{eom:xi}) 
at late times $k \tau \to 0$.
A decaying bound state is given by
\begin{equation}
 u_d=A_d \tau^2 \,, \quad
 \chi_d= C_d \,\tau^2 \sqrt{\sinh(Hz)}
 \left[ 1+\frac{\cosh(Hz)}{2} \log \left(\frac{\cosh(Hz)-1}{\cosh(Hz)+1}\right)\right]\,,
\label{boundd}
\end{equation}
where $C_d$ is linearly related to $A_d$. We will re-derive these solutions later 
on in a slightly different language. For the time being we do not need explicit relations
between the coefficients entering (\ref{boundg}) or (\ref{boundd}).

On the other hand, the continuum modes are
\begin{equation}
u_{\lambda} = g_{\lambda} \tau^{\alpha_{\lambda}}, \quad
\chi_{\lambda} = f_{\lambda}(z) \tau^{\alpha_{\lambda}},
\end{equation}
where
\begin{equation}
\alpha_{\lambda}= \frac{1}{2} \pm i\lambda \, ,
\end{equation}
and in the high energy regime
\begin{equation}
 f_{\lambda}(z) =N_{\lambda} \cos(\lambda z + \beta_{\lambda}).
\end{equation}
The coupled equations determine $g_{\lambda}$ and $\beta_{\lambda}$ in terms of
$N_\lambda$. Again, we do not need the explicit relations here and in what follows.
It suffices to say that using these solutions, one can check that 
the Wronskian between
the bound states and the continuum modes vanishes \footnote{This is 
actually obvious from the fact that the continuum modes oscillate in
$\tau$, while the bound states do not. If non-zero, the Wronskian
of a bound state and a continuum mode would oscillate in time, which
would contradict its constancy.} as $\tau \to 0$, and hence
at all $\tau$.
In other words, they are orthogonal to each other. Furthermore, 
unlike the growing bound state solution, the continuum modes decay away
as $\tau \to 0$, so they do not have any effect at super-horizon scales.
Thus, we focus on the bound states in the rest of this paper.

We are interested in the final amplitude of perturbations, which
is determined by the the growing bound state.
On large scales, the growing mode  gives
\begin{equation}
{\cal R}_c = -\frac{H^2}{\dot{\phi}} A_g \, .
\label{rc}
\end{equation}
The amplitude $A_g$ has to be determined by  quantisation
of the inflaton and metric fields on sub-horizon scales.
We thus have to evolve the growing bound state backwards in time.
For reasons that will become clear later, the early-time
behaviour of the decaying bound state is also of importance.

\section{Bound-state solutions}
\label{sec:boundstate}
\subsection{Numerical analysis}
\label{sec3a}

We now investigate what becomes of the bound states at early times.
By using the numerical method developed in Ref.~\cite{HK}, the coupled equations
are evolved backwards in time using the late time bound state solutions,
Eqs.~(\ref{boundg}) or (\ref{boundd}), as initial conditions.
In this section, we rescale to dimensionless variables
\begin{equation}
\label{rescale}
 \tau \to k^{-1} \tau \,,\qquad z \to H^{-1} z \,, \qquad
\xi \to \kappa_5^2 \dot{\phi} k^{-1} \xi,
\end{equation}
to simplify the problem further. Recall 
 that the evolution backwards 
in time is the evolution forward in $\tau$ because we redefined the time 
variable $\tau \to - \tau$.
The boundary equations (\ref{eom:xi}) and~(\ref{junction:xi}) become
\begin{eqnarray}
\xi_{,\tau \tau} + \xi &=& - 2 \beta^2 \frac{\chi_{_{\rm b}}}{\tau},
 \nonumber\\
\left[ \chi' +\frac{1}{2} \frac{N'}{N} \chi \right]_{_{\rm b}}&=& \tau \xi.
\end{eqnarray}
In terms of rescaled $\xi$ and $\chi$, the Wronskian is
\begin{equation}
W(f^*,f)= \frac{k}{(\kappa_5^2 \dot{\phi})^2 }
\bar{W}(f^*,f), \quad \bar{W}(f^*,f) = W_{\xi}(\xi^*,\xi) 
+ W_{\chi}(\chi^*,\chi),
\end{equation}
where
\begin{equation}
W_{\xi}(\xi^*,\xi) = \xi^*\xi_{,\tau}-\xi \xi_{,\tau}^*,\quad
W_{\chi}(\chi^*,\chi)= 2 \beta^2 \int^\infty_{z_{_{\rm b}}} dz\,
\big(\chi^* \chi_{,\tau} - \chi \chi_{,\tau}^*\big) \,.
\label{wronski}
\end{equation}
Fig.~1 shows the behaviour of $\chi(z,\tau)$ for the mode that
grows as $\tau \to 0$, i.e., has late-time asymptotics
(\ref{boundg}). It is clear that
the bound state persists at large $\tau$. As can be seen from Fig.~2,
the amplitude of $\chi$ on the brane increases for large
$\tau$ like $\chi(z_b,\tau) \propto \tau^{1/3}$. As for the
profile in the bulk, $\chi(z,\tau)$ is more and more concentrated near
the brane at
earlier times (larger $\tau$). This means that the solution for $\chi(z, \tau)$
is not a separable function of $\tau$ and $z$.
On the other hand, the amplitude of $\xi$ approaches a
constant for large $\tau$ as is seen from Fig.~3. 
The right panel of Fig.~3 shows
that the phase difference between the growing mode and
the decaying mode is roughly equal to $\pi/2$.
Fig.~4 shows the behaviour of the Wronskian between the growing mode 
and the decaying mode.
For large $\tau$ the Wronskian is conserved independently 
for $W_{\xi}$ and $W_{\chi}$. In these calculations, we have set 
$A_g=A_d =1$. 

\begin{figure}[h]
\label{3d}
 \begin{center}
\includegraphics[width=12cm]{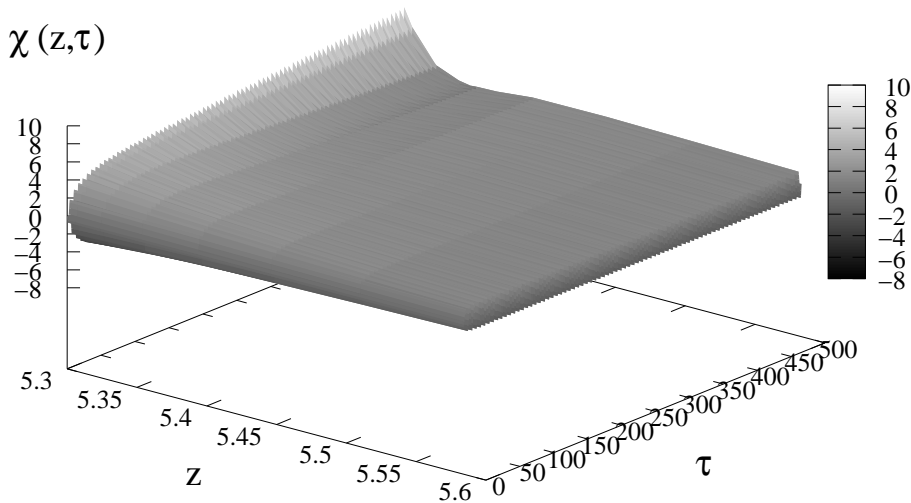}
\caption[]{The growing mode solution for $\chi(z,\tau)$. We take $\beta=0.1$
and $H/\mu=100$.
The brane is located at $z_b=5.3$. }
\end{center}
\end{figure}

\begin{figure}[h]
\label{figchi}
 \begin{center}
\includegraphics[width=18cm]{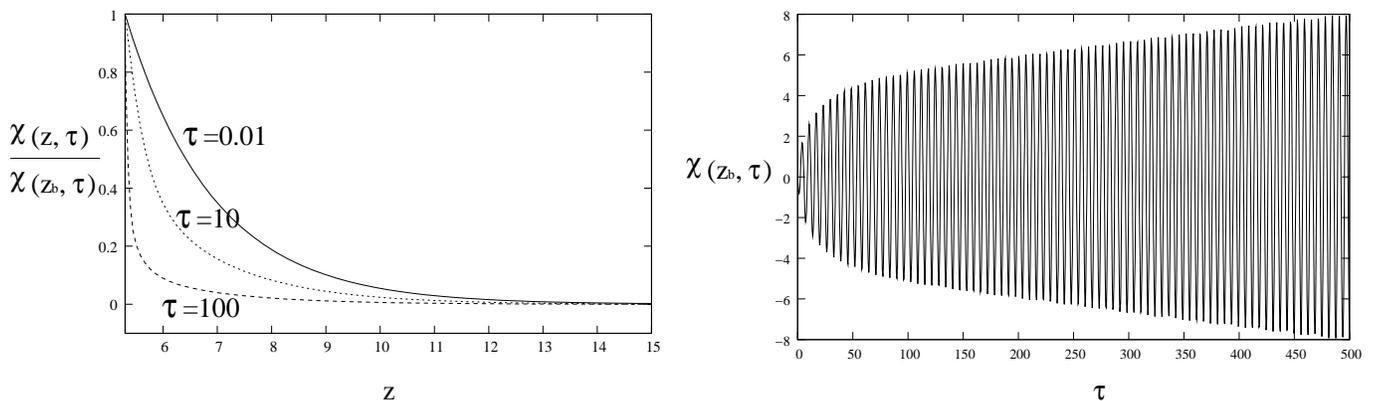}
\caption[]{The growing mode solution for $\chi(z,\tau)$
for the same parameters as in Fig.~1. Left: the profile of
$\chi(z,\tau)$ in the bulk. The amplitude is normalised so that the
amplitude on the brane is equal to 1. Right: behaviour of $\chi$ on the
brane, $\chi(z_b,\tau) $.
}
\end{center}
\end{figure}

\begin{figure}[http]
\label{figxi}
 \begin{center}
\includegraphics[width=18cm]{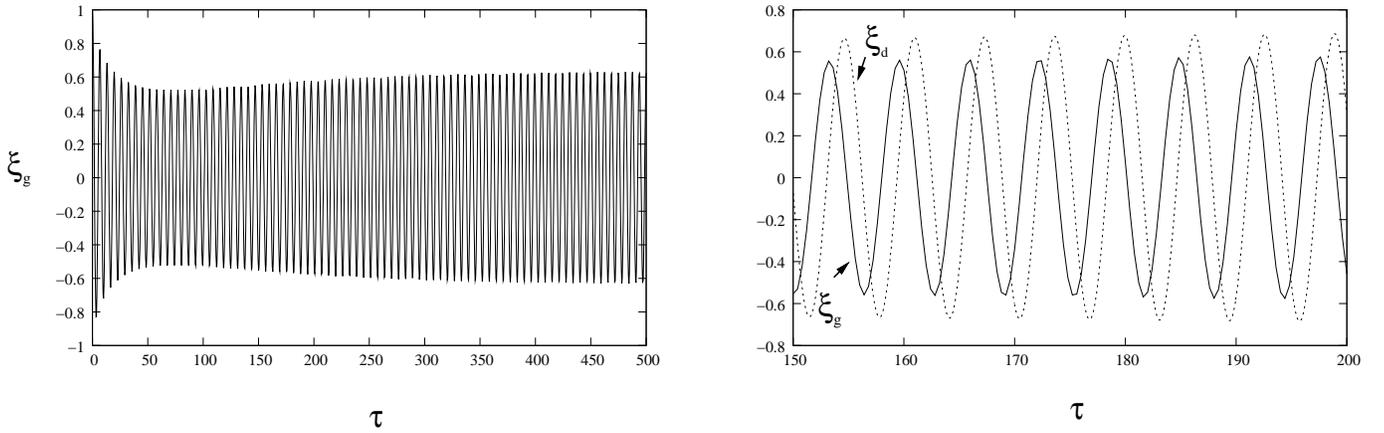}
\caption[]{Left: The growing mode solution for $\xi(\tau)$
for the same parameters as in Fig.~1. Right:
Comparison between the growing mode  $\xi_g$ (solid line)
and  decaying mode
$\xi_d$ (dotted line).}
\end{center}
\end{figure}

\begin{figure}[h]
\label{Wronskian}
 \begin{center}
\includegraphics[width=8cm]{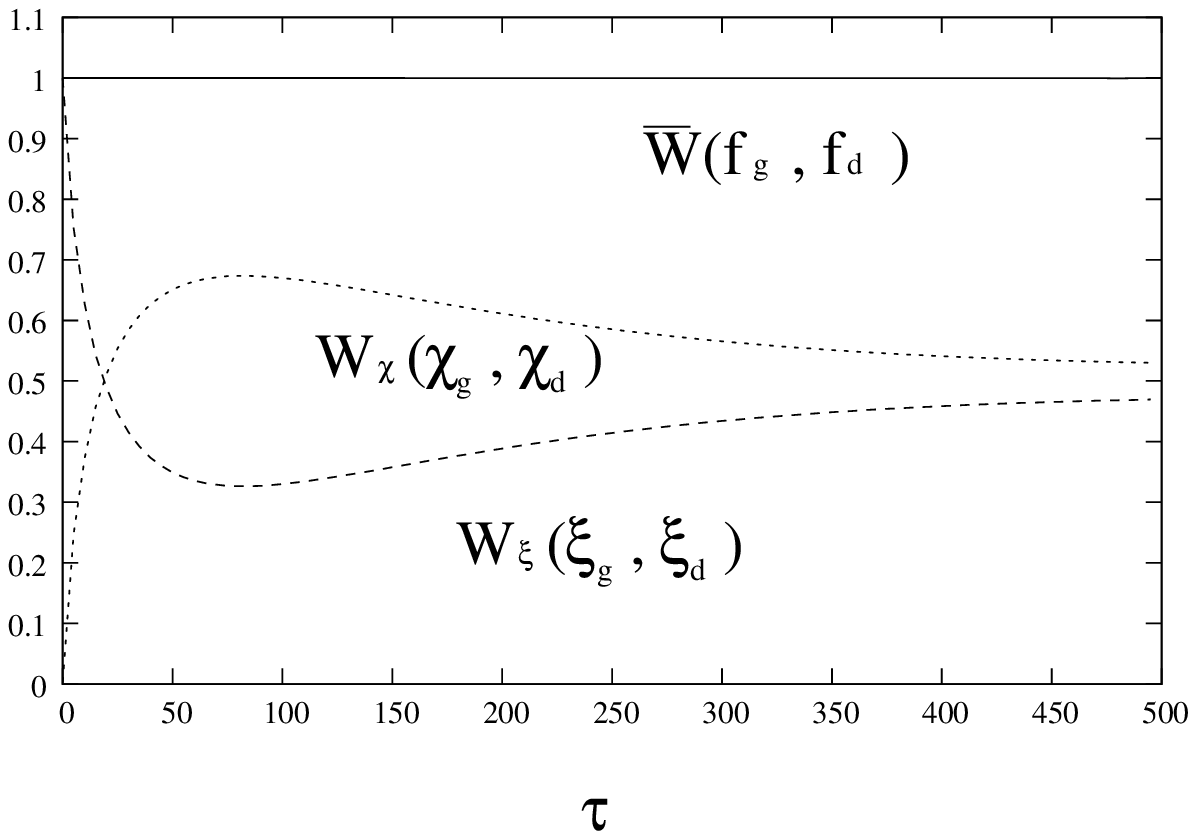}
\caption[]{Behaviour of the Wronskian between the growing 
 and  decaying bound states, $\bar{W}(f_g,f_d)$.}
\end{center}
\end{figure}

\subsection{Kaluza--Klein tower}
The numerical results indicate that even at early times, there exists
the bound states. In Refs.~\cite{KLMW,KMW}, the behaviour of the bound states
has been analyzed by performing the perturbative expansion in  $\beta^2$.
It was shown that, in terms of the Kaluza--Klein decomposition,
the bound state is composed of an infinite ladder of 
modes  with negative mass squared. Due to the excitation of the infinite ladder
of the modes, the naive perturbation expansion in $\beta^2$ breaks down
at early times~\cite{KMW}. In this paper, we derive the solutions
without performing the perturbative expansion in $\beta^2$.

The bulk equation of motion for $\chi$, Eq.~(\ref{eom:chi}) 
is straightforward to solve.
The solutions
can be written in the form~\cite{KLMW}
\begin{equation}
 \label{sum:chi}
 \chi=\int d\nu \, B_\nu f_\nu(z) T_\nu(\tau) \,,
\end{equation}
where $\nu$ is related to the conventional Kaluza--Klein mass as
\begin{equation}
\nu = \sqrt{\frac{9}{4}- \frac{m^2}{H^2}},
\end{equation}
$\tau^{-1/2} T_\nu(\tau)$ is in general a linear combination of the Bessel
function $J_\nu(\tau)$ and Neumann function $N_\nu(\tau)$, and $(\sinh
z)^{-1/2} f_\nu(z)$ is a linear combination of the Legendre functions
$P_{\nu-1/2}(\cosh z)$ and $Q_{\nu-1/2}(\cosh z)$.  
We are interested in the bound states for which $\chi$ vanishes as $\tau \to 0$
and/or $z \to \infty$, so we consider the solutions composed of the
following mode functions:
\begin{equation}
 T_\nu(\tau)= \tau^{1/2} J_\nu(\tau) \,, \qquad
 f_\nu(z) = \left( \frac{\sinh z}{\sinh z_{_{\rm b}}}\right)^{1/2}
 \frac{Q_{\nu-1/2} (\cosh z)}{Q_{\nu-1/2} (\cosh z_{_{\rm b}})} \,,
\label{modedef*}
\end{equation}
where $\mbox{Re} ~ \nu > 0$. Our normalisation ensures $f_\nu(z_{_{\rm b}})=1$.
 We can then write
the brane field $\xi$ in terms of the mode functions $T_\nu(\tau)$ as
\begin{equation}
 \label{sum:xi}
 \xi = \int d\nu \, C_\nu T_\nu(\tau) \,.
\end{equation}

The brane equation of motion (\ref{eom:xi}) and the junction
condition (\ref{junction:xi}) then reduce to two coupled difference
equations relating the coefficients $B_\nu$ and $C_\nu$ entering
Eqs.~(\ref{sum:chi}) and~(\ref{sum:xi}),
\begin{gather}
\label{Rec1}
 \frac{F_{\nu+1}}{2(\nu+1)} B_{\nu+1} + \frac{F_{\nu-1}}{2(\nu-1)} B_{\nu-1} = C_\nu \,,\\
\label{Rec2}
 \frac{(\nu+1)^2-1/4}{2(\nu+1)} C_{\nu+1} + \frac{(\nu-1)^2-1/4}{2(\nu-1)} C_{\nu-1} = -2\beta^2 B_\nu \,,
\end{gather}
where we use the shorthand
\begin{equation}
 F_\nu = \frac{Q^1_{\nu-1/2}(\cosh z_{_{\rm b}})}{Q_{\nu-1/2}(\cosh z_{_{\rm b}})} \; .
\end{equation}
We now derive two independent bound state solutions to these coupled equations.

To get a bound state we require that $B_\nu=0$ whenever
$\mbox{Re}~\nu \le 0$.  It can be seen from Eqs.~(\ref{Rec1}) and
(\ref{Rec2}) that there are two such solutions: one for which 
$C_{2n-1/2}$ and $B_{2n+1/2}$ are non-vanishing only, and the other with
non-zero values only of $C_{2n+1/2}$ and $B_{2n+3/2}$, where $n = 0,1,2,\dots$.  
Importantly, we choose $C_{-1/2}$ real, then all $C_{2n-1/2}$ and $B_{2n+1/2}$
obtained from Eqs.~(\ref{Rec1}) and (\ref{Rec2}) are real; the same is true for
the second solution.

At late times, the modes with higher $n$
decay faster, thus the late time solutions are described by the lowest
mode:
\begin{eqnarray}
\chi_{g} &=& B_{1/2} f_{1/2}(z) T_{1/2}(\tau), \quad
\xi_{g} = C_{-1/2} T_{-1/2}(\tau), \nonumber\\
\chi_{d} &=& B_{3/2} f_{3/2}(z) T_{3/2}(\tau), \quad
\xi_{d} = C_{1/2} T_{1/2}(\tau) \; ,
\label{largemode}
\end{eqnarray}
where $C_{-1/2}$ and $C_{1/2}$ are arbitrary constants and according to
(\ref{Rec1})
\begin{equation}
   B_{1/2} = \frac{C_{-1/2}}{F_{1/2}} \; , \quad
  B_{3/2} = \frac{3 C_{1/2}}{F_{3/2}} \; .
\end{equation}
We see that the solution that starts
from $C_{-1/2}$ corresponds to the growing mode 
(\ref{boundg}) with $A_g =C_{-1/2} \sqrt{2/\pi}$, while
 the solution that starts from $C_{1/2}$ corresponds to the decaying mode
(\ref{boundd}) with $A_d = C_{1/2} \sqrt{2/\pi}$. Hence, solving the recurrence
equations (\ref{Rec1}) and (\ref{Rec2}) 
starting from $C_{-1/2}$ or $C_{1/2}$ is equivalent to evolving
the late time bound state solutions backwards
to obtain their early time behaviour.
We have solved the recurrence relations numerically, the results are shown in
Figs.~5 and 6. Using these solutions, we have
checked that the solutions $\chi(\tau, z)$, $\xi (\tau)$ obtained by summing
up the ladder of modes are in excellent agreement with the results of
numerical simulations described in section~\ref{sec3a}.

\begin{figure}[h]
 \begin{center}
\includegraphics[width=9cm]{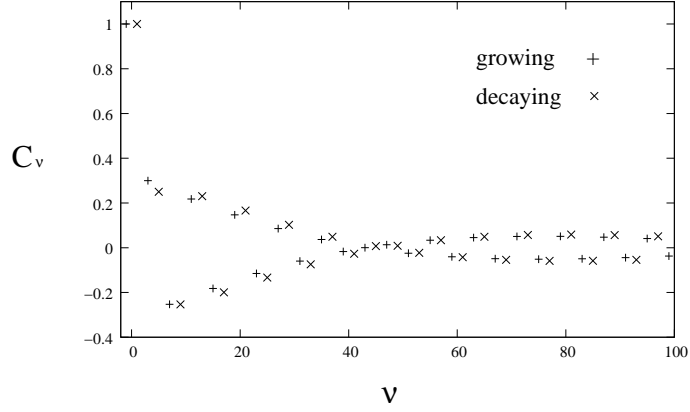}
\caption[]{Solutions for $C_{\nu}$ with $\beta^2 =0.1$. The growing 
mode solution with $C_{2n-1/2}$ is shown by $+$ and the 
decaying mode solution with $C_{2n+1/2}$ is shown by $\times$.  We took $C_{1/2}=C_{-1/2}=1.$}
\end{center}
\end{figure}

\begin{figure}[h]
 \begin{center}
\includegraphics[width=18cm]{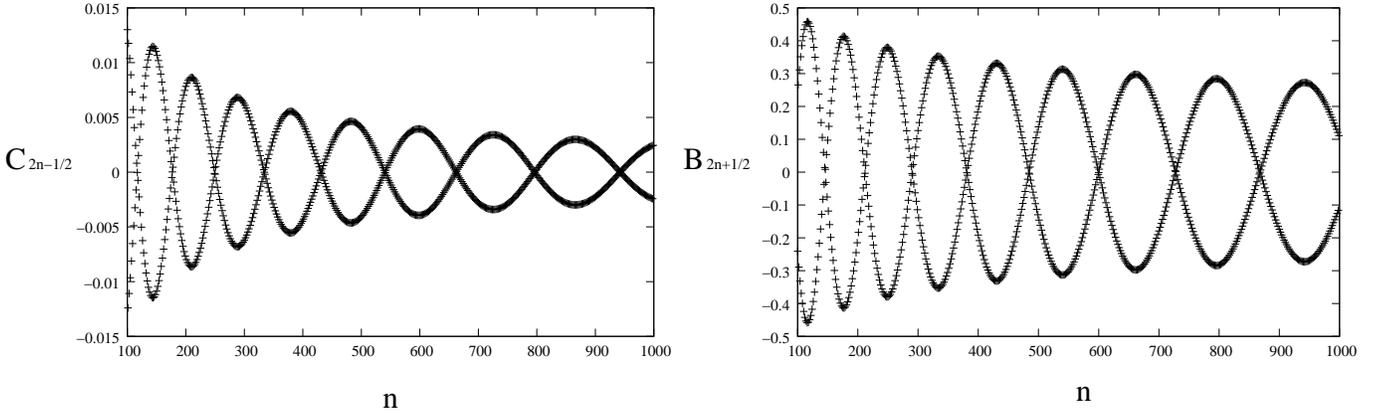}
\caption[]{Solutions for $C_{2n-1/2}$ and $B_{2n+1/2}$ for large $n$. }
\end{center}
\end{figure}

\subsection{High energy limit: analytical treatment}

{}From now on we will work in the high energy limit $H/\mu \gg 1$.  
In this limit one has
\begin{equation}
F_\nu \to  -\left(\nu +\frac12\right) \quad \text{as} \quad H/\mu \to \infty \,.
\label{asymF}
\end{equation}
and the expression (\ref{modedef*}) for the mode function simplifies to
\begin{equation}
f_\nu (z) = e^{-\nu (z-z_b)}.
\end{equation}
In order to derive analytically
the asymptotic behaviour of the bound state solutions at early times, 
$\tau \to \infty$,
it is useful to have
an asymptotic expansion for $B_{\nu}$ and $C_{\nu}$ for large $n$ or
equivalently for large $\nu$.  {}From the numerical solutions we find
 that the sequences $B_{2n+1/2}$ and $C_{2n-1/2}$ alternate in
sign (and the same for the other solution) so we define new sequences
\begin{equation}
 \overline{B}_{2n+1/2}=(-1)^{n+1} B_{2n+1/2} \,, \quad  
 \overline{C}_{2n-1/2} = (-1)^n C_{2n-1/2} \, ,
\end{equation}
for the growing bound state and 
\begin{equation}
 \overline{B}_{2n+3/2}=(-1)^{n+1} B_{2n+3/2} \,, \quad 
 \overline{C}_{2n+1/2} = (-1)^n C_{2n+1/2} \, ,
\end{equation}
for the decaying bound state.
For large $\nu$, we can write the difference equations as
\begin{eqnarray}
 \overline{C}_\nu &=& \left[ \frac{x-1/2}{2x} \overline{B}_x\right]_{x=\nu-1}^{x=\nu+1}
 \simeq \frac{d}{d\nu} \left(\overline{B}_\nu \right) \,,
\label{sec3.14}\\
 -2\beta^2 \overline{B}_\nu &=& \left[ \frac{x^2-1/4}{2x} \overline{C}_x \right]_{x=\nu-1}^{x=\nu+1}
 \simeq \frac{d}{d\nu} \left( \nu \overline{C}_\nu \right) \,,
\label{sec3.15}
\end{eqnarray}
provided that  $\overline{B}_\nu$ and $\overline{C}_\nu$ are smooth functions of $\nu$
whose derivatives vanish as $\nu \to \infty$.
{}From Eqs.~(\ref{sec3.14}) and (\ref{sec3.15}) it follows that  
the coefficients $\overline{B}_\nu$  satisfy the differential equation
\begin{equation}
 \frac{d}{d\nu} \left( \nu \frac{d\overline{B}_\nu}{d\nu} \right) + 2\beta^2 \overline{B}_\nu=0\,,
\end{equation}
to the leading order in $\nu^{-1}$, which has solutions
\begin{equation}
\label{asympBbar}
 \overline{B}_\nu \propto Z_0 \big( 2 \beta \sqrt{ 2 \nu} \big) \,,
\end{equation}
where $Z_0$ is a linear combination of the 
Bessel function of order 0,
\begin{equation}
Z_{0,i} \big(2 \beta\sqrt{2 \nu}\big)= A_{J,i}
J_{0}\big(2 \beta\sqrt{2 \nu}\big) + A_{N,i}
N_{0}\big(2 \beta\sqrt{2 \nu}\big), \label{Zsol}
\end{equation}
where $i=g$ for the growing mode solution and $i=d$ for the decaying
mode solution, and $A_{J,i}$ and $A_{N,i}$ are yet undetermined real constants,
two for the growing mode and two for the 
decaying mode. The reality of  $A_{J,i}$ and $A_{N,i}$
follows from the reality of $B_\nu$.
From (\ref{Rec1}) we find, again to the  leading order in $\nu^{-1}$,
\begin{equation}
 \overline{C}_\nu = \frac{d\overline{B}_\nu}{d\nu} = -\beta \sqrt\frac{2}{\nu}\,
 Z_1 \big( 2 \beta \sqrt{2 \nu} \big) \, ,
\end{equation}
where $Z_1$ is the combination of $J_1$ and $N_1$ with the same
 coefficients $A_{J,i}$ and $A_{N,i}$ as in Eq.~(\ref{Zsol}). 
In the large-$n$ asymptotics, the coefficients for the growing
bound-state solution are thus
\begin{equation}
\label{tower2}
 B_{2n+1/2} = (-1)^{n+1} Z_0 \big( 4 \beta \sqrt{n} \big) \,,\qquad
 C_{2n-1/2} = (-1)^{n+1} \beta \,\frac{Z_1 \big( 4 \beta \sqrt{n} \big)}{\sqrt
{n}} \,.
\end{equation}
In the same way, the decaying bound state has
\begin{equation}
\label{tower}
 B_{2n+3/2} = (-1)^{n+1} Z_0 \big( 4 \beta \sqrt{n} \big) \,,\qquad
 C_{2n+1/2} = (-1)^{n+1} \beta \,\frac{Z_1 \big( 4 \beta \sqrt{n} \big)}{\sqrt
{n}} \,,
\end{equation}
We have checked that the agreement between these asymptotic solutions
and the numerical solutions of Eqs.~(\ref{Rec1}) and~(\ref{Rec2}) is
excellent after we fit the two coefficients $A_{J,i}$ and $A_{N,i}$ for each mode.


To the zeroth order in $\beta^2$, the constants $A_{J,i}$ and
$A_{N,i}$ can be expressed analytically through $C_{-1/2}$ (for the growing mode)
and $C_{1/2}$ (for the decaying mode). Since for the growing mode
$C_{\nu}$ is $O(\beta^2)$
for $\nu > - 1/2$, we can neglect $C_{\nu}$ with $\nu > -
1/2$ in Eq.~(\ref{Rec1}) for the this mode. Likewise, we can neglect
$C_{\nu}$ with $\nu > 1/2$ for the decaying mode.
Then, as was first shown in~\cite{KLMW},
Eq.~(\ref{Rec1}) can be solved as 
\begin{equation}
B_{2n+1/2} = (-1)^n \left(2 n+\frac{1}{2} \right)
\frac{2}{F_{2n+1/2}} C_{-1/2}, \quad B_{2n+3/2} = (-1)^n \left(2
n+\frac{3}{2} \right) \frac{2}{F_{2n+3/2}} C_{1/2}.
\end{equation}
Making use of the expression
(\ref{asymF}) valid in the high energy regime, we get 
for $n \gg 1$ 
\begin{equation}
B_{2n+1/2} = (-1)^{n+1} 2 C_{-1/2}, \quad B_{2n +3/2} = (-1)^{n+1} 2
C_{1/2}. \label{pert}
\end{equation}
Comparing with the non-perturbative solutions (\ref{tower}) and
(\ref{tower2}), we find that the perturbative solutions (\ref{pert}) are valid
for $\beta \sqrt{n} \ll 1$. Taking the small argument limit of
Eqs.~(\ref{tower}) and (\ref{tower2}), we obtain
\begin{equation}
A_{J,g} = 2 C_{-1/2},\quad  A_{J,d} = 2 C_{1/2}, \quad
A_{N,g}=A_{N_d}=0, \label{solbc}
\end{equation}
at the zeroth order in $\beta^2$. The latter relations effectively provide
matching between the late-time behaviour (\ref{largemode}) and the early-time
asymptotics which we are about to find. This matching, however, is valid
to the zeroth order in $\beta^2$ only.

It should be emphasized that the
perturbative approach breaks down for $\beta \sqrt{n} \gtrsim 1$. This
is because, even if each $C_{\nu}$ with $\nu \neq \pm 1/2$ is
$O(\beta^2)$, the contribution to $B_{\nu}$ from $C_{\nu}$ in
Eq.~(\ref{Rec1}) is $O(n \times C_{\nu}) \sim O(n \beta^2)$. 
As we will see later, large-$n$ modes become dominant for large $\tau$.
Then even for small $\beta^2$, the naive perturbation theory in 
$\beta$ breaks down at large $\tau$.
This is in accord with Ref.~\cite{KMW}.
For finite $\beta^2$, $A_{J,i}$ and $A_{N,j}$ are found from 
numerical solutions of Eqs.~(\ref{Rec1}) and (\ref{Rec2})
and they are shown in Fig.~7.

\begin{figure}[h]
\label{AJN}
 \begin{center}
\includegraphics[width=18cm]{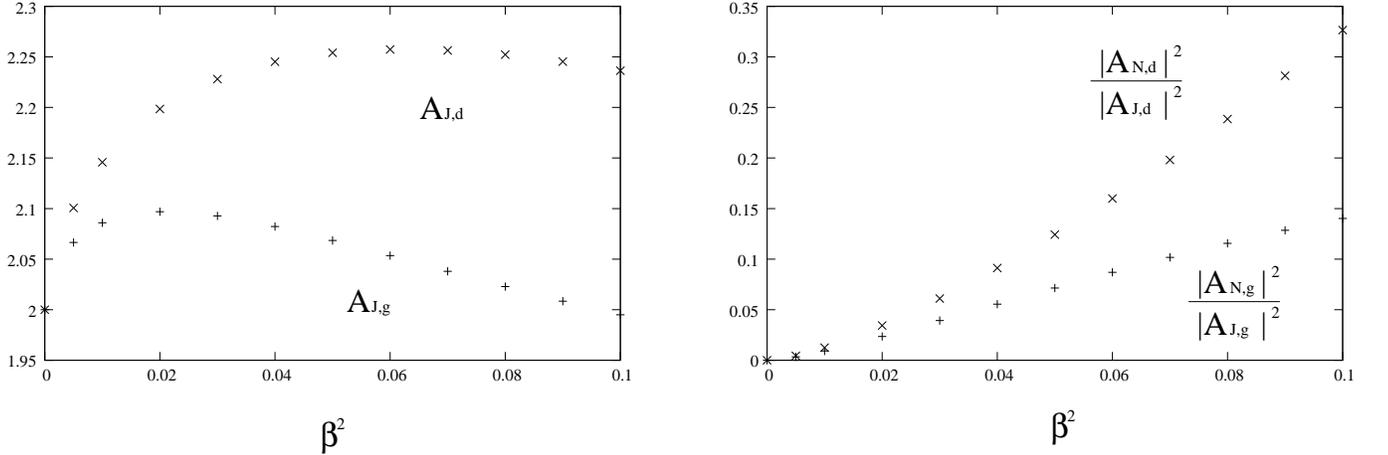}
\caption[]{Numerical solutions for the 
coefficients $A_{J,i}$ and $|A_{N,i}|^2/|A_{J,i}|^2$, defined in
  Eq.~(\ref{Zsol}), for different
  values of $\beta^2$. We took $C_{1/2}=C_{-1/2}=1.$}
\end{center}
\end{figure}

Using the asymptotic expressions for the coefficients given in
Eqs.~(\ref{tower}) and (\ref{tower2}), 
we can approximate the solutions at $z-z_b \ll 1$ by
\begin{eqnarray}
 \label{early_chi_tower}
\chi_g(\tau,z) &\simeq& \sum_n (-1)^{n+1} Z_0\big(4\beta\sqrt{n}\big) \sqrt\tau J_{2n+1/2}
(\tau) e^{- 2 n (z-z_b)},
\nonumber\\
\label{early_chi-tower2}
\chi_d(\tau,z) &\simeq& \sum_n (-1)^{n+1} Z_0\big(4\beta\sqrt{n}\big) \sqrt\tau J_{2n+3/2}
(\tau) e^{- 2 n (z-z_b)} \, .
\end{eqnarray}
For large $\tau$,  the sums over the modes can be evaluated explicitly.
First we use the asymptotic expression for the Bessel functions,
\begin{equation}
 J_\nu(\tau) \simeq \sqrt\frac{2}{\pi\nu\tan\theta} \cos \big( \nu\tan\theta - \nu\theta -\pi/4 \big),
\end{equation}
where $\cos\theta = \nu/\tau$, see Ref.~\cite{GradRyz}.  The sums are
dominated by the modes for which $n/\tau \ll 1$ but $n^2/\tau \gg 1$. In this regime
the asymptotic expression simplifies to
\begin{equation}
 J_{n+1/2}(\tau) =  \sqrt\frac{2}{\pi\tau}\cos\big(\tau + n^2/(2 \tau) - n \pi/2 
 -\pi/2 \big)  \,.
\end{equation}
The standard large-argument asymptotic form~\cite{GradRyz} for the
Bessel function $Z_0$ in Eq.~(\ref{tower}) enables one to approximate the sums in
Eqs.~(\ref{early_chi_tower}) and (\ref{early_chi-tower2}) as
\begin{eqnarray}
\label{chi_integral}
 \chi_g &\propto&
 (A_{J,g} + i A_{N,g}) \int dn \,n^{-1/4} \, \exp\left(- 2 n (z-z_b) 
 - 4i\beta\sqrt{n} + 
i\tau + 2in^2 \tau^{-1} \right)+c.c.\, , \nonumber\\
 \chi_d &\propto&
 (A_{J,d} + i A_{N,d}) \int dn \,n^{-1/4} 
\, \exp\left(- 2 n (z-z_b) 
  - 4i\beta\sqrt{n} + i\tau + 2in^2 \tau^{-1} - \frac{\pi}{2}i \right)+c.c. \, ,
\end{eqnarray}
where we keep only the terms that give dominant contributions to the integrals
(see below).
The phase difference of $\pi/2$ between the growing mode  and
decaying mode  solutions comes
from the fact that the growing mode and decaying mode solutions contain $J_{2n+1/2}$
and $J_{2n+3/2}$, respectively.
At large $\tau$ (early times) these integrals are saturated
by modes with large $n$. This means, in particular, that we cannot
truncate the ladder of modes if we wish to describe the early time
behaviour~\cite{KMW}.
The leading-order contribution comes from the saddle point at
\begin{equation}
 n_{_{\rm saddle}} = \frac{1}{2^{2/3}} \beta^{2/3} \tau^{2/3} -
 \frac{i}{3} (z-z_b) \tau \,.
\end{equation}
The integrals in Eq.~(\ref{chi_integral}) can thus be approximated as
\begin{eqnarray}
 \chi_g &\propto& (A_{J,g}+i A_{N,g}) \tau^{1/3} \exp \left( i\tau-\frac{3i}{2^{1/3}} \beta^{4/3} \tau^{1/3} - 2^{1/3}\beta^{2/3}\tau^{2/3} 
(z-z_b)
 +\frac{i}{3} \tau (z-z_b)^2 \right) +c.c. \, ,\nonumber\\
 \chi_d &\propto& (A_{J,d}+i A_{N,d}) \tau^{1/3} \exp \left( i\tau-\frac{3i}{2^{1/3}} \beta^{4/3} \tau^{1/3} - 2^{1/3}\beta^{2/3}\tau^{2/3} 
(z-z_b)
 +\frac{i}{3} \tau (z-z_b)^2 - \frac{\pi}{2} i \right) +c.c. \, .
\label{asymptochi}
\end{eqnarray}
Note that at large $\tau$, these wave functions 
are localized in a narrow region in the bulk,
$ z-z_b \sim (\beta \tau)^{-3/2} \ll 1$, as we anticipated
when writing Eq.~(\ref{early_chi_tower}).
The terms omitted in Eq.~(\ref{chi_integral}) do not have saddle points near
the semi-axis of positive $n$ and thus give the contributions which are exponentially
suppressed as compared to Eq.~(\ref{asymptochi}).
In the same way we can find the asymptotic behaviour of $\xi$,
\bea
\xi_{g} &\propto& (A_{J,g}+i A_{N,g}) \exp
\left(i \tau -\frac{3i}{2^{1/3}} \beta^{4/3} \tau^{1/3}\right) +c.c \, ,\nonumber\\
\xi_{d} &\propto& (A_{J,d}+i A_{N,d}) \exp
\left(i \tau  -\frac{3i}{2^{1/3}} \beta^{4/3} \tau^{1/3} - \frac{\pi}{2}i \right) + c.c \, .
\label{asymptoxi}
\eea

These asymptotic expressions nicely explain the behaviour of solutions
at early times ($\tau \gg 1$) obtained in our numerical simulations.
We  see that the amplitude of $\chi(z_b,\tau)$ increases
as $\chi(z_b, \tau)  \propto \tau^{1/3}$ while the amplitude of
$\xi$ tends to a constant at large $\tau$. At the same time,
$\chi(z,\tau)$ is increasingly dominated by large-$n$ modes and becomes
more and more closely bound to the brane at large $\tau$. This behaviour
ensures that the Wronskian for $\chi$, given in Eq.~(\ref{wronski}),
is conserved at late times. We also notice that the phase difference
between the growing mode and  decaying mode solutions
is approximately equal to $\pi/2$, provided that $A_{J,i} \gg A_{N,i}$. 
As we saw above, the latter inequality is indeed valid at small $\beta^2$,
so the pattern shown in the right panel of Fig.~3 also has 
its analytical explanation.

To conclude this section, we notice that at early times, the original
coupled system Eqs.~(\ref{eom:chi}),
(\ref{eom:xi}) and (\ref{junction:xi}) may be treated within the WKB 
approximation. This is done in Appendix~A. The WKB analysis is in prefect
agreement with the results Eqs.~(\ref{asymptochi}) and (\ref{asymptoxi}).

\section{The quantum amplitude}
In the preceding section, we have found the early time behaviour of the bound
state solutions, Eqs.~(\ref{asymptochi}) and (\ref{asymptoxi}). For finite $\beta^2$
they are parameterized by the real coefficients $A_{J,i}$ and $A_{N,i}$
where $i=g,d$ for the growing and decaying bound state solution, 
respectively.
These coefficients can be found numerically, see Fig.~7. Importantly,
the expressions (\ref{asymptochi}) and (\ref{asymptoxi}) show that
despite the complicated  behaviour of the bound states at early
times, the division of solutions into positive- and negative-frequency
parts is well defined as $\tau \to \infty$. This is also clear from the
WKB analysis, see Appendix~A. Thus,
 as in the familiar four-dimensional case, we can straightforwardly
quantise the perturbations and 
define the vacuum state on small scales (early times)  to
determine the final amplitude of the inflaton fluctuations.  We follow
the method developed in Refs~\cite{wronskian1, wronskian2} to relate
the amplitude of the growing mode solution to the creation and
annihilation operators. In principle, this method involves the construction
of properly normalised positive- and negative-frequency solutions at early
times and the evaluation of the inner product (Wronskian) between these
solutions and the bound states. However, there is a short-cut: all we need
are the values of the real coefficients  $A_{J,i}$ and $A_{N,i}$. We first
explain this short-cut by making use of the familiar four-dimensional
example, and then proceed to the brane-world model.
%
In this section, we restore $k$, but we are 
still working with redefined time $\tau \to -\tau$.

\subsection{Four-dimensional case}
Let us begin with the standard four-dimensional case.
At the first order in the slow-roll parameters, the equation for
$u$ is 
\begin{equation}
u_{,\tau \tau} + k^2 u - \frac{1}{\tau^2}(2 + 6 \epsilon_{_H} -3 \eta_{_H})u =0.
\end{equation}
In the same way as in Eq.~(\ref{def:xi}), we  define
\begin{equation}
\xi = u_{,\tau} + (1 + 2 \epsilon_{_H} - \eta_{_H}) \frac{u}{\tau}.
\label{xiu}
\end{equation}
The equation for $\xi$ is 
\begin{equation}
\xi_{,\tau \tau}+ k^2 \xi -\frac{2 \epsilon_{_H} - \eta_{_H}}{\tau^2} \xi =0,
\label{xi}
\end{equation}
The Wronskian is defined as
\begin{equation}
W(\xi^*,\xi) = \frac{1}{k^2} (\xi^* \xi_{,\tau} - \xi \xi^*_{,\tau}).
\end{equation}
Neglecting $k^2$ in Eq.~(\ref{xi}), one obtains 
the growing mode  and the decaying mode solutions at late times, $\tau \to 0$,
\begin{equation}
\xi_g \to  (k\tau)^{1/2-\nu} 2^{\nu} \frac{1}{\Gamma(- \nu+1)},
\quad
\xi_d \to (k\tau)^{1/2+\nu} 2^{-\nu} \frac{1}{\Gamma(\nu+1)}.
\label{latexi}
\end{equation}
where $\nu=1/2+2 \epsilon_{_H} - \eta_{_H}$ and we included numerical factors for later
convenience.
%
We are interested in  quantum field which behaves at late times as
\begin{equation}
\hat{\xi} \to \hat{Z} \xi_g,
\end{equation}
where $\hat{Z}$ is a time-independent quantum
operator. This field is quantised on small scales where we can expand
$\hat{\xi}$ in terms of negative- and positive-frequency modes,
\begin{equation}
\hat{\xi} = \hat{a} \varphi^{(-)} + \hat{a}^{\dagger} \varphi^{(+)},
\end{equation}
where $\hat{a}$ and $\hat{a}^{\dagger}$ are annihilation and
creation operators, respectively, which define the vacuum
\begin{equation}
\hat{a} |0 \rangle =0.
\end{equation}
We should note that due to the redefinition of time $\tau \to - \tau$, 
the definitions of the negative- and positive-frequency
functions are 
$\varphi^{(-)} \propto e^{ik\tau}$ and $\varphi^{(+)} \propto 
e^{-i k \tau}$.
The mode functions should be normalised as
\begin{equation}
W(\varphi^{(-)},\varphi^{(+)}) = -i \; .
\label{wtrivial}
\end{equation}
The latter condition  ensures the canonical commutational relation between $\xi$
and its conjugate momentum.

Our aim is to express $\hat{Z}$ in terms of $\hat{a}$ and $\hat{a}^{\dagger}$.
We exploit the constancy of the Wronskian for this purpose.
At late times, $\hat{Z}$ can be related to the Wronskians,
\begin{equation}
\hat{Z} = \frac{W(\hat{\xi},\xi_d)}{W(\xi_g,\xi_d)}.
\end{equation}
Here $W(\xi_g,\xi_d)$ can be calculated using the late time
solutions (\ref{latexi}) for $\xi_g$ and $\xi_d$,
\begin{equation}
W(\xi_g, \xi_d) = \frac{1}{k} \left(\frac{2}{\pi}\right) \sin(\nu \pi).
\label{late}
\end{equation}
On the other hand, $W(\hat{\xi},\xi_d)$ can be calculated
at early times. For this purpose, we expand the growing mode 
and the decaying mode solutions in $\varphi^{(+)}$ and $\varphi^{(-)}$
at early times,
\begin{equation}
\xi_g = c_g \varphi^{(-)} + c_g^{*} \varphi^{(+)},
\quad
\xi_d = c_d \varphi^{(-)} + c_d^{*} \varphi^{(+)}.
\label{genexpansion}
\end{equation}
Then, $W(\hat{\xi},\xi_d)$ is calculated as
\begin{equation}
W(\hat{\xi}, \xi_d) = -i \left(\hat{a} c_d^{*} - \hat{a}^{\dagger} c_d \right).
\end{equation}
As usual, the operator $\hat{Z}$ corresponds to the Gaussian random field, 
fully characterised by 
the vacuum expectation value of its square. The latter is evaluated as
\begin{equation}
\langle\hat{Z}^{\dagger} \hat{Z} \rangle = k^2 \left(\frac{\pi}{2} \right)^2
|c_d|^2 (\sin \nu \pi)^{-2}.
\label{fin4d}
\end{equation}
Thus the problem reduces to finding $c_d$, which is the expansion coefficient
for the decaying mode solution.

Now, the trick is that we do not need to know the precise expressions for
$\varphi^{(\pm)}$ to determine $|c_d|$.
Indeed, by solving for the growing mode and decaying mode
 backwards, it is possible, at least in principle (and in practice
in the four-dimensional example), to derive the early-time
solutions for $\xi_g$, $\xi_d$. In the four-dimensional case they are
\begin{equation}
\xi_g = N_g \cos(k \tau +\Delta_g), \quad
\xi_d = N_d \cos(k \tau + \Delta_d).
\label{xiN}
\end{equation}
This immediately impies that the positive- and negative-frequency modes have the form
\begin{equation}
\varphi^{(\pm)} = |\varphi| e^{\mp i \delta} e^{\mp i k \tau} \, ,
\end{equation}
where the normalisation factor $|\varphi|$ may be found from (\ref{wtrivial}),
but will not be needed in what follows, and $\delta$ is some phase (which is
irrelevant in the four-dimensional case).
Comparing Eqs.~(\ref{genexpansion}) and (\ref{xiN}), we get
\begin{equation}
\frac{|c_d|}{|c_g|} = \frac{N_d}{N_g}, \quad 
\mbox{Arg} \, c_g - \mbox{Arg}\, c_d \equiv \delta_g- \delta_d
=\Delta_g - \Delta_d.
\label{eqc1}
\end{equation}
The additional information
on $c_d, c_g$ and $|\delta_d - \delta_g|$ can be obtained from
the Wronskian $W(\xi_g, \xi_d)$.
In terms of $c_g$ and $c_d$, the Wronskian is calculated
as
\begin{equation}
W(\xi_g, \xi_d) = 2 |c_g| |c_d| |\sin (\delta_g-\delta_d)| \, .
\label{eqc2}
\end{equation}
Now we can use  Eq.~(\ref{late}) together with Eqs~(\ref{eqc1}) and (\ref{eqc2})
to obtain $|c_d|^2$ in terms of $N_d/N_g$ and $|\Delta_g - \Delta_d|$,
\begin{equation}
|c_d|^2 = \frac{1}{2} \frac{N_d}{N_g} \frac{1}{|\sin (\Delta_g-\Delta_d) |}
\frac{1}{k} \left(\frac{2}{\pi} \right) \sin \nu \pi.
\label{cd-4d}
\end{equation}
Thus, the quantity $|c_d|^2$ entering (\ref{fin4d}) is expressed in terms of
the ratio $N_d/N_g$ of the amplitudes and difference $(\Delta_g - \Delta_d)$
of the phases of the decaying and growing modes evolved back in time.

In the four-dimensional case,
the growing mode and the decaying mode solutions are
\begin{equation}
\xi_g = (k \tau)^{1/2} J_{-\nu}(k \tau),
\quad
\xi_d = (k \tau)^{1/2} J_{\nu}(k \tau),
\end{equation}
At late times, $\tau \to 0$, they have the form (\ref{latexi}), while
at early times, $k\tau \gg 1$, the solutions become
\begin{equation}
\xi_g \to \sqrt{\frac{2}{\pi}} \cos \left(k \tau+ \frac{1}{2} \nu \pi -
\frac{1}{4} \pi \right),
\quad
\xi_d \to \sqrt{\frac{2}{\pi}} \cos \left(k \tau- \frac{1}{2} \nu \pi -
\frac{1}{4} \pi \right).
\label{earlyxi}
\end{equation}
We thus find
\begin{equation}
\frac{N_d}{N_g}=1, \quad \Delta_g -\Delta_d = \nu \pi.
\end{equation}
Then $|c_d|^2$ is obtained as
\begin{equation}
|c_d|^2 = \frac{1}{2k} \left(\frac{2}{\pi} \right),
\label{cd4d}
\end{equation}
and the expectation value for $\hat{Z}$ squared is given by
\begin{equation}
\langle \hat{Z}^{\dagger}\hat{Z} \rangle = \frac{k \pi}{4} (\sin \nu \pi)^{-2}.
\label{zz}
\end{equation}
In this way, from the solutions for the growing and decaying modes,
we can determine the quantum amplitude of the fluctuations.
We note in parenthesis that the slow-roll corrections do not affect
the value of $|c_d|$ in the four-dimensional case. This is
because the slow-roll effects do not play
any role at early times in Eq.~(\ref{xi}).

The quantity that is related to observables is the
comoving curvature perturbation,
\begin{equation}
{\cal R}_c = - \frac{H}{a \dot{\phi}} u.
\end{equation}
Using the result for the quantum expectation value for $\hat{Z}^\dagger \hat{Z}$, 
Eq.~(\ref{zz}),
and the late-time asymptotics (\ref{latexi}), we get from (\ref{xiu}) at 
$k\tau \ll 1$
%
\be
\langle u^2 \rangle^{1/2} \to 2^{2 \epsilon_{_H} - \eta_{_H}}
\frac{\Gamma (3/2+2 \epsilon_{_H}-\eta_{_H})}{\Gamma (3/2)}
\frac{1}{\sqrt{2k}} (k \tau)^{-1-2 \epsilon_{_H}+\eta_{_H}}.
\label{asym_sl_Munov_vl}
\ee
Expanding the Gamma function in Eq.~(\ref{asym_sl_Munov_vl}), we find
\be
\langle u^2 \rangle^{1/2} \to
\Big\{1+ (2\epsilon_{_H}-\eta_{_H}) (2- \gamma - \ln 2)
\Big\} \frac{1}{\sqrt{2k}}(k\tau)^{-1-2\epsilon_{_H}+\eta_{_H}} \, .
\label{uk_int_of_ah}
\ee
Here we have used the formula for the psi function
\be
\psi(3/2) \equiv \frac{\Gamma'(3/2)}{\Gamma (3/2)}
= 2  -\gamma - 2 \ln 2,
\ee
where $\gamma$ is the Euler constant.
The power spectrum of the curvature perturbation is then given by
\be
{\cal{P}}_{{\cal R}_c} (k) = \frac{k^3}{2 \pi^2}
\langle{\cal R}_c^2 \rangle = \frac{k^3}{2 \pi^2} \left(\frac{H}{a \dot{\phi}}\right)^2
\langle u^2 \rangle.
\label{def_curvpert}
\ee
Using the fact that conformal time is given, up to the first order
in slow-roll parameters, by
\be
\tau = \frac{1}{aH} (1+\epsilon_{_H}),
\ee
one finds the power spectrum of the curvature perturbation, 
\bea
{\cal{P}}_{{\cal R}_c} &=&  \left(\frac{H}{\dot{\phi}}\right)^2
\left(\frac{H}{2 \pi}\right)^2\Big[1 + 2 S(\epsilon_{_H}, \eta_{_H}) \Big],
 \quad S(\epsilon_{_H}, \eta_{_H})=
(2 \epsilon_{_H} -\eta_{_H})(2 - \ln 2 - \gamma) - \epsilon_{_H} \, ,
\label{s_l_correction}
\eea
where $S$ is known as the Stewart--Lyth correction~\cite{SL}.

\subsection{Brane-world case}
We now apply the method presented in the preceding subsection to the
brane-world case. The late time behaviour of $\xi$ is the same as in the
four-dimensional case: at late times we have
\be
\hat{\xi} \to \hat{Z} \xi_g.
\ee
We expand the quantum operators $\hat\xi$ and $\hat\chi$ at early times as
\bea
\hat\xi &=& \hat{a} \varphi^{(-)} + \hat{a}^{\dagger} \varphi^{(+)},
 \nonumber\\
\hat\chi &=& \hat{a} \phi^{(-)} + \hat{a}^{\dagger} \phi^{(+)},
\label{expand}
\eea
where $\varphi^{(\pm)}$ and $\phi^{(\pm)}$ are the
positive and negative frequency solutions to
the coupled equations, obeying the normalisation condition
\be
W(f^{(-)}, f^{(+)})
=-i,
\ee
where $f^{(\pm)}=(\varphi^{(\pm)}, \phi^{(\pm)})$, and the Wronskian
has precisely the form (\ref{wr33}).
This normalisation ensures the canonical quantisation conditions
for the fields, as is shown in Appendix B.

The growing mode solution and the decaying mode solution
are also expanded as
\begin{equation}
\xi_{i} \to c_i \varphi^{(-)} + c_i^{*} \varphi^{(+)}, \quad
\chi_{i} \to c_i \phi^{(-)} + c_i^{*} \phi^{(+)},
\end{equation}
where $i=g,d$. At late times, the contribution from
$\chi$ to the Wronskian can be neglected and the dominant contribution
comes from the lowest mode in the ladder of the modes, Eq.~(\ref{largemode}),
\begin{equation}
\xi_g = (k \tau) J_{-1/2}(k \tau), \quad
\xi_d = (k \tau) J_{1/2}(k \tau),
\end{equation}
where we set $C_{1/2}=C_{-1/2}=1$.
In this way we obtain the Wronskian,
\begin{equation}
W(f_g,f_d) =  \frac{1}{k} \frac{2}{\pi}.
\end{equation}
Then, as in the four-dimensional case, the expectation value of 
the quantum operator $\hat{Z}^\dagger \hat{Z}$ is calculated as
\begin{equation}
\langle \hat{Z}^{\dagger} \hat{Z} \rangle = k^2 \left(\frac{\pi}{2} \right)^2
|c_d|^2.
\end{equation}
The virtue of the approach outlined above is that it works even in cases
when the early-time solutions have complicated form. In the brane-world case,
the positive- and negative-frequency solutions can be read off from 
Eqs.~(\ref{asymptochi}) and (\ref{asymptoxi}), and they indeed are rather
complicated. Nevertheless, the above argument goes through, and 
%
we find, in exact analogy with the result (\ref{cd-4d}),
\begin{equation}
|c_d|^2 = \frac{1}{2} \frac{N_d}{N_g} \frac{1}{|\sin (\Delta_g-\Delta_d)|}
\frac{1}{k} \left(\frac{2}{\pi} \right),
\end{equation}
%
where
\begin{equation}
\frac{|N_d|^2}{|N_g|^2} = \frac{A_{J,d}^2+A_{N,d}^2}{A_{J,g}^2+A_{N,g}^2},\quad
\Delta_g-\Delta_d = \frac{\pi}{2} +\left[\arctan \frac{A_{N,g}}{A_{J,g}} -
\arctan \frac{A_{N,d}}{A_{J,d}} \right].
\label{NDel}
\end{equation}
As a cross check, let us show that the standard four-dimensional result
is recovered at the zeroth order in $\beta^2$. At this order, 
the solutions for $A_{J,i}$ and $A_{N,i}$
are given by Eq.~(\ref{solbc}). Then Eq.~(\ref{NDel}) yields
\begin{equation}
\frac{|N_d|^2}{|N_g|^2}=1, \quad \Delta_g-\Delta_d= \frac{\pi}{2},
\end{equation}
and we get
\begin{equation}
|c_d|^2 = \frac{1}{2k} \left(\frac{2}{\pi} \right),
\end{equation}
which agrees with the four-dimensional result, Eq.~(\ref{cd4d}).

For finite $\beta^2$ we use numerical solutions for
$A_{J,i}$ and $A_{N,i}$ shown in Fig.~7 to evaluate
$N_d/N_g$ and $(\Delta_g-\Delta_d)$. The results are
presented in Fig.~8. Using these numerical results, we obtain
the expectation
value 
of interest,
\be
\langle \hat{Z}^{\dagger} \hat{Z} \rangle
=\frac{k \pi}{4} \Big[ 1 + K(\beta^2) \Big] \, ,
\ee
where $K(\beta^2)$ is shown in Fig.~9. As is seen from Fig.~9, 
$K(\beta^2) \sim O(\beta^2) \sim O(\epsilon_{_H})$.

\begin{figure}[h]
 \begin{center}
 \includegraphics[width=18cm]{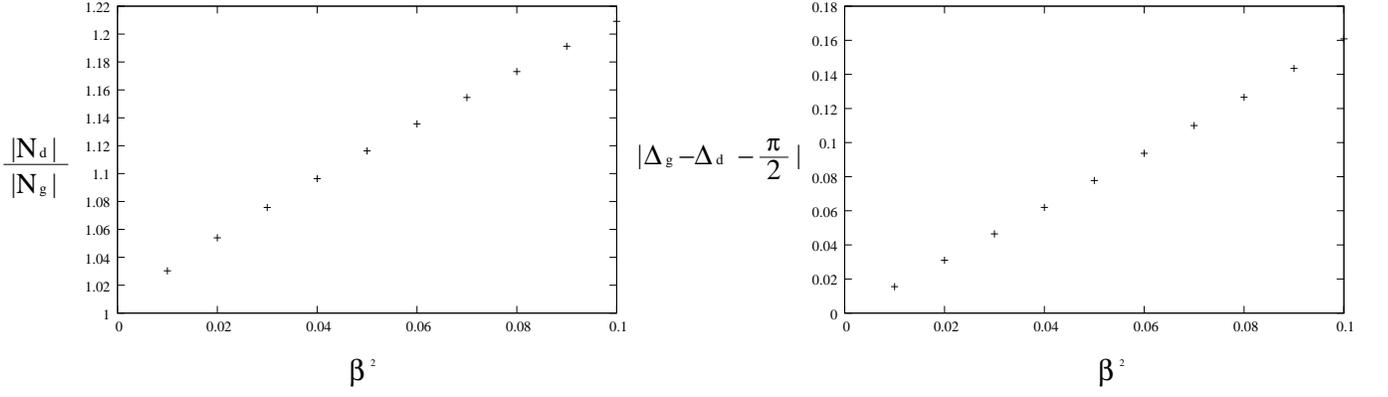}
\caption[]{Numerical solutions for 
$|N_d|/|N_g|$ and $|\Delta_g-\Delta_d - \pi/2|$.}
\end{center}
\label{amplitude}
\end{figure}


\begin{figure}[h]
 \begin{center}
 \label{result}
\includegraphics[width=9cm]{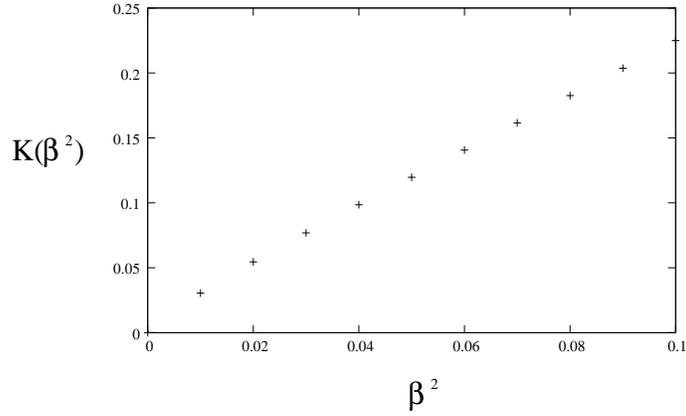}
\caption[]{Numerical solutions for $K(\beta^2)$.}
\end{center}
\end{figure}

\subsection{Amplitude of curvature perturbations}
Finally, the amplitude of the comoving curvature perturbations
is calculated using Eqs.~(\ref{rc}) and (\ref{xig}).
At late times, $\hat{\xi}$ behaves as
\begin{equation}
\hat{\xi} = \hat{Z} \sqrt{\frac{2}{\pi}}.
\end{equation}
Then, comparing this with Eq.~(\ref{xig}),
we find the amplitude of $A_g$,
\begin{equation}
\langle A_g^2 \rangle=\frac{1}{k^4} \frac{2}{\pi}
\langle \hat{Z}^{\dagger} \hat{Z} \rangle
= \frac{1}{2k^3} \Big[1+K(\beta^2) \Big].
\end{equation}
Using Eq.~(\ref{xig}), the spectrum of the curvature
perturbations is evaluated as
\bea
{\cal{P}}_{{\cal R}_c} &=&  \left(\frac{H}{\dot{\phi}}\right)^2
\left(\frac{H}{2 \pi}\right)^2 \Big[ 1 + K(\beta^2) \Big].
\label{kbeta}
\eea
We should note that we did not take into account the standard
slow-roll corrections at the super-horizon scales in this
calculation. The correction $K(\beta^2)$ is solely 
due to the coupling between the inflaton and the bulk
metric perturbations. Although the mode functions are significantly
affected by this coupling at early times, the final correction to the quantum
amplitude at late times is suppressed by the slow-roll parameter even
in the high energy regime.

As the correction
$K(\beta^2)$ is  of the same order as the usual slow-roll corrections,
we should also take into account the corrections to the amplitude
at the superhorizon scales. Since $K(\beta^2)$ is already the
first-order correction, the slow-roll corrections to $K(\beta^2)$
are of higher order. Then we just need to add the Stewart--Lyth
correction. Indeed, neglecting the coupling to the
bulk metric perturbations that is already first order
in the slow-roll parameter, the equation for $u$ is given by
\begin{equation}
u_{,\tau \tau} + k^2 u - \frac{1}{\tau^2}(2 + 6 \epsilon_{_H} -3 \eta_{_H})u =0.
\end{equation}
We find that this is exactly the same as in the four-dimensional case. Thus,
the Stewart--Lyth correction is precisely the same as well; this 
was first shown in Ref.~\cite{RL}.
Then at the first order in slow-roll, we get the
final result
\bea
{\cal{P}}_{{\cal R}_c} &=&  \left(\frac{H}{\dot{\phi}}\right)^2
\left(\frac{H}{2 \pi}\right)^2
\Big[ 1 + K(\beta^2) + 2S(\epsilon_{_H}, \eta_{_H}) \Big] \, ,
\label{correction}
\eea
where $S=(2 \epsilon_{_H} -\eta_{_H})(2 - \ln 2 - \gamma) - \epsilon_{_H}$
is the Stewart--Lyth correction. Fig.~10 shows the complete first order 
corrections in the brane-world model at high energies,  $H \gg \mu$,
compared with the 
Stewart--Lyth corrections.

\begin{figure}[h]
 \begin{center}
\includegraphics[width=16cm]{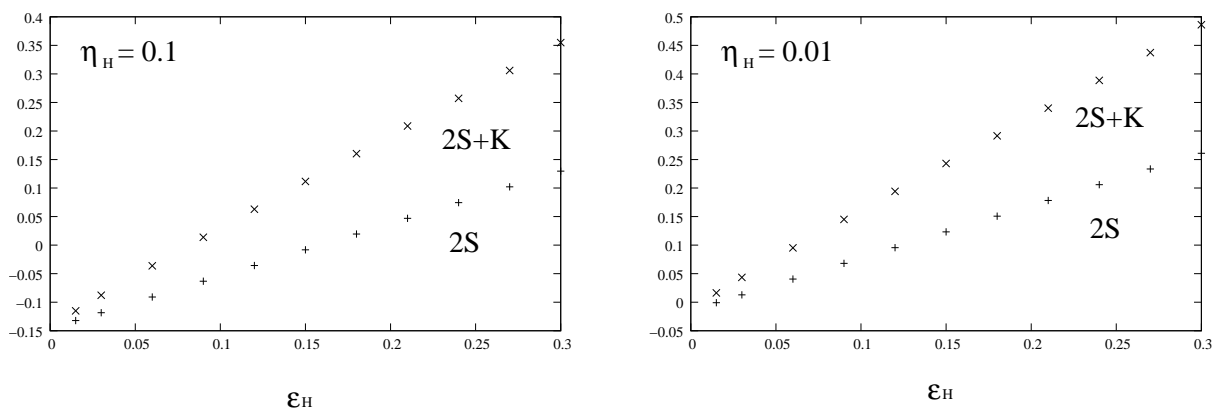}
\caption[]{First order corrections $2 S(\epsilon_{_H}, \eta_{_H}) +K(\beta^2)$
in the high energy regime $H \gg \mu$,
compared with $2 S(\epsilon_{_H}, \eta_{_H})$ for $\eta_{_H}=0.1$ (left) and 
$\eta_{_H}=0.01$ (right).}
\end{center}
\end{figure}

\section{Conclusions}
In this paper, we performed the quantisation of the
inflaton perturbations during slow-roll inflation driven by an
inflaton confined to a brane in the Randall-Sundrum
brane world-model. The slow-roll dynamics 
leads to a finite coupling between the inflaton fluctuations
on the brane and metric perturbations in the bulk.
We first studied the effect of this coupling at the first
order in the slow-roll expansion by keeping the coupling
term while neglecting all other slow-roll corrections.
This enabled us to deal with five-dimensional perturbations
around the de Sitter brane for which the bulk wave equation is separable
and solutions can be described as sums of bulk modes.
At late times there exists  bound states with single bulk
modes, and the growing bound state determines the final
amplitude of the comoving curvature perturbation on large scales. The
continuum modes were shown to be orthogonal to the bound state in the
sense that the Wronskian between them vanishes;
furthermore, the continuum modes decay away at late times. 
Thus we could neglect
the continuum modes in our analysis.

We evolved this late time bound state backwards numerically to find its
early time behaviour. It was shown that the behaviour of the inflaton
fluctuations is significantly affected by its coupling to an infinite
tower of bulk modes, even at the first order in the slow-roll
approximation. Asymptotically, the amplitude of $\xi$, that describes
the inflaton perturbations on the brane, becomes constant while the
amplitude of $\chi$, that describes the bulk metric perturbations,
increases on the brane at early times. It was also found that $\chi$
is more and more concentrated near the brane at early times, becoming
dominated by modes with large (and negative) Kaluza--Klein
mass-squared. We found that it is possible to reproduce these
numerical solutions by the summation of the infinite ladders of
modes. The growing mode  and decaying mode solutions have
different ladders of modes, which leads to a phase shift between them.
The infinite sum of the modes can be evaluated using a saddle point
method and analytic formula for the asymptotic solutions was derived.

We then performed the quantisation of the bound state.
We developed a scheme where the quantum amplitude
of the perturbations can be derived by finding the
early time behaviour of the growing mode and the
decaying mode.
In this method, the constancy of the
Wronskian is exploited to connect the quantum
operator at late times to the creation and annihilation
operators defined at early times, without the explicit
analysis of the early-time
positive- and negative-frequency functions and the 
calculation of the Bogoliubov coefficients. This method was
applied to the standard four-dimensional case to re-derive
the Stewart--Lyth correction, which arises from corrections to
the field evolution after horizon exit.
We then applied this method to our brane-world model.
In this case, there is a correction that arises from the
modified solutions for the perturbations on small scales.
Thus, this correction is due to the coupling of the inflaton 
to the
bulk metric perturbations. Despite the fact that
the behaviour of perturbations on small scales is qualitatively
different due to this coupling, it was found that particle
production is suppressed by the smallness of the
coupling. The correction to the final amplitude
was thus shown to be of the order of the 
slow-roll parameter, $\epsilon_{_H}$,
even in the high energy regime.

Because the effect of the mixing between the brane inflaton
fluctuations and the bulk metric perturbations is of the 
first order in the slow-roll parameter, 
the complete first order corrections
consist of the correction  $K(\beta^2)$ coming from the brane-bulk mixing
and the Stewart--Lyth correction $S(\epsilon_{_H}, \eta_{_H})$, which was shown
to have the same form as in the four-dimensional case. Fig.~10 is our 
main result showing the complete first order corrections in the 
brane-world model.

In four-dimensional general relativity the  corrections to
the evolution of inflaton fluctuations that arise at the first order in
the
slow-roll parameters can be neglected for wavelengths much smaller
than the Hubble scale. Thus, the vacuum fluctuations on sub-horizon
 scales are essentially the same as those of a massless field in
the de Sitter space-time~\cite{Mukhanov}. The only change in the amplitude
of quantum fluctuations comes from the modification of
 evolution on super-horizon
scales, leading to the familiar Stewart--Lyth correction~\cite{SL}.

By contrast, in the brane world the coupling between the inflaton
fluctuations and bulk metric perturbations that arises at the first order
in the slow-roll parameters leads to a radically different picture of the
bound sates at early times compared with the zeroth-order
solution. Short wavelength inflaton fluctuations do not decouple
in the high energy regime ($H\gg\mu$) from the bulk metric perturbations
and support an
infinite tower of bulk modes, which become increasingly dominated by
large-$n$ modes on small scales. Nonetheless, the mixing between
positive- and negative-frequency modes is absent in the sub-horizon regime,
while additional mixing near the horizon crossing
is suppressed by slow-roll parameters, so the final correction
to the amplitude of fluctuations on super-horizon scales is
again first order in the slow-roll parameters.

\section*{Acknowledgments}
KK is supported by PPARC and AM by PPARC grant PPA/G/S/2002/00576.
VR is supported in part by
RFBR grant 05-02-17363a. TH is supported by JSPS.
This collaboration resulted from the meeting and workshop
\emph{Brane-World Gravity: Progress and Problems} held in at the
University of Portsmouth in September 2006.

\appendix
\section{WKB approximation for early-time solution}
Let us consider early times, i.e., large $\tau$. In this case 
we can find the bound state solutions to the coupled system of equations
(\ref{eom:chi}),
(\ref{eom:xi}) and (\ref{junction:xi})
by making use of
the WKB approximation.  Using the rescaling given 
in Eq.~(\ref{rescale}) we approximate the solutions by
\begin{equation}
 \xi = A(\tau) \exp (i S_\xi(\tau)) \,, \qquad
 \chi = B(\tau,z) \exp(i S_\chi(\tau,z)) \,,
\end{equation}
where $A$ and $B$ are slowly varying functions but the exponential factors
are  varying rapidly.
Consistency with the brane equation of motion (\ref{eom:xi}) requires that
\begin{equation}
 S_\xi(\tau) = S_\chi (\tau, z_{_{\rm b}}) = S_{_{\rm b}} (\tau)\,.
\end{equation}
In the WKB regime and in the high-energy limit,
 Eqs.~(\ref{eom:xi}) and (\ref{junction:xi}) reduce to
\begin{gather}\left( 
 -\left(\frac{\partial S_{_{\rm b}}}{\partial \tau}
 \right)^2 +1 \right) A = -\frac{2\beta^2}{\tau} B_{_{\rm b}} \,,\\
 i \left[\frac{\partial S_\chi}{\partial z}\right]_{_{\rm b}} B_{_{\rm b}} = \tau A \,.
\end{gather}
From these equations we obtain the  boundary condition for $S_\chi$,
\begin{equation}
\left(
 -\left(\frac{\partial S_{_{\rm b}}}{\partial \tau}
 \right)^2 +1 \right)
 \left[\frac{\partial S_\chi}{\partial z}\right]_{_{\rm b}} 
= 2 i \beta^2 \,.
\end{equation}
Let us also write the 
equation of motion for $\chi$, Eq.~(\ref{eom:chi}), at the boundary,
\begin{equation}
-\left(\frac{\partial S_{_{\rm b}}}{\partial \tau}
 \right)^2 +1  + \frac{1}{\tau^2} 
\left[\frac{\partial S_\chi}{\partial z}\right]^2_{_{\rm b}} = 0 \,.
\end{equation}
The latter two equations give
\begin{equation}
\label{S_condition1}
\left(\frac{\partial S_{_{\rm b}}}{\partial \tau}
 \right)^2   = 1 - \left( \frac{2\beta^2}{\tau}\right)^{2/3} \,, \qquad
 \left[\frac{\partial S_\chi}{\partial z}\right]_{_{\rm b}} 
= i 2^{1/3} \beta^{2/3} \tau ^{2/3} \,.
\end{equation}
One can use this approach 
to solve for $S_\chi$ as a Taylor series about the brane position.  The next derivative is
\begin{equation}
\label{S_condition2}
 \left[\frac{\partial^2 S_\chi}{\partial z^2}\right]_{_{\rm b}} = \frac{2}{3} \tau \,,
\end{equation}
and so on.  Note that this solution indeed
satisfies the applicability conditions for the WKB approximation.

We can obtain the amplitudes $A$ and $B$ in a similar way by
using the next order of  the WKB approximation, finding that
\begin{equation}
 A = \text{const} \,,\qquad B_{_{\rm b}} = \text{const} \cdot \tau^{1/3} \,.
\end{equation}
Note that the amplitude of $\xi$ is 
constant but the amplitude of $\chi$ increases like $\tau^{1/3}$.  
The expansion near the brane gives
\begin{equation}
 \chi(\tau,z) \propto \tau^{1/3} \, e^{- 2^{1/3} \beta^{2/3} \tau^{2/3} 
 (z-z_b)} \,
e^{-iS_{_{\rm b}}} \, .
\label{wkbchi}
\end{equation}
This expression implies, in particular,
that the contribution from the small-$z$ region to the normalisation integral
\begin{equation}
 \int dx \,(\chi^* \chi_{,\tau} - c.c.),
\end{equation}
is constant in $\tau$.
The asymptotic solution Eq.~(\ref{wkbchi}) agrees with the solution obtained
in Section~III.B, Eq.~(\ref{asymptochi}).

\section{quantisation condition for coupled system}
We normalised the positive- and negative-frequency modes as
\be
W(f^{(-)}, f^{(+)})
= \frac{1}{k^2} (
\varphi^{(-)} \varphi_{,\tau} ^{(+)}- \varphi^{(+)} \varphi_{,\tau}
^{(-)}) + \frac{1}{3 \kappa_5^2}
\int_{z_b}^{\infty} dz (
\phi^{(-)} \phi^{(+)}_{,\tau}- \phi^{(+)} \phi^{(-)}_{,\tau})
=-i.
\label{wrapp}
\ee
We justify this as follows. In terms of $\xi$ and $\chi$, the quadratic action is given by
\begin{eqnarray}
S&=&\frac{1}{2 k^2} \int d \tau (\xi_{,\tau}^2 - k^2 \xi^2)
+\frac{1}{6 \kappa_5^2} \int^{\infty}_{z_b}
dz d\tau \left(
\chi_{,\tau}^2 - k^2 \chi^2 - \frac{\chi'^2 +(U(z)-2H^2) \chi^2}{(H \tau)^2}
\right) \nonumber\\
&-& \frac{\dot{\phi}}{3H} \int d \tau \frac{1}{\tau} \chi \xi
+ \frac{1}{12 \kappa_5^2} \left( \frac{N'}{N} \right)_{_{\rm b}} \int d \tau
\frac{\chi^2}{(H \tau)^2}.
\label{acapp}
\end{eqnarray}
Therefore, the conjugate momenta are
\begin{equation}
\pi_{\xi} = -\frac{\xi_{,\tau}}{k^2}, \quad
\pi_{\chi} = -\frac{1}{3 \kappa_5^2} \chi_{,\tau},
\end{equation}
where the minus sign comes 
from the redefinition of time $\tau \to -\tau$.
From the expansions of the fields
in terms of $\hat{a}$ and $\hat{a}^{\dagger}$,
\bea
\xi &=& \hat{a} \varphi^{(-)} + \hat{a}^{\dagger} \varphi^{(+)}, \nonumber\\
\chi &=& \hat{a} \phi^{(-)} + \hat{a}^{\dagger} \phi^{(+)},
\label{expand2+}
\eea
we can express $\hat{a}$ and $\hat{a}^{\dagger}$ as
\begin{equation}
W(f^{(-)}, f) =- i \hat{a}^{\dagger} , \quad
W(f^{(+)},f) = i \hat{a}.
\end{equation}
Then the commutation relation between $\hat{a}$ and $\hat{a}^{\dagger}$
is calculated as
\bea
[\hat{a}, \hat{a}^{\dagger}] &=& [W(f^{(+)},f), W(f^{(-)},f)] \nonumber\\
&=& \frac{1}{k^2} (\varphi^{(-)} \varphi_{,\tau}^{(+)} -
\varphi^{(+)} \varphi_{,\tau}^{(-)} )[\xi, \pi_{\xi}]
+ \frac{1}{3 \kappa_5^2} \int dz (\phi^{(-)} \phi^{(+)}_{,\tau}
-\phi^{(+)} \phi^{(-)}_{,\tau}) [\chi, \pi_{\chi}] \nonumber\\
&+& \frac{1}{k^2} \int dz (\varphi_{,\tau}^{(+)} \phi^{(-)}
- \varphi_{,\tau}^{(-)} \phi^{(+)}) [\xi, \pi_{\chi}]
+ \frac{1}{3 \kappa_5^2} \int dz
(\phi_{,\tau}^{(+)} \varphi^{(-)} - \phi_{,\tau}^{(-)} \varphi^{(+)})
[\chi, \pi_{\xi}].
\eea
Imposing the canonical commutational relations
\begin{equation}
[\xi, \pi_{\xi}] =i ,\quad[\chi (z), \pi_{\chi} (z^\prime)]=i \delta(z,z^\prime),
\quad [\xi, \pi_{\chi}]=0, \quad
[\chi, \pi_{\xi}]=0,
\end{equation}
and using the normalisation (\ref{wrapp}) 
we get
\begin{equation}
[\hat{a}, \hat{a}^{\dagger}]=1.
\label{aa}
\end{equation}
This justifies our normalisation.

A cross check is the calculation of the Hamiltonian which is derived
from the action (\ref{acapp}) in the standard way. By making use of
the decomposition (\ref{expand2+}), the Hamiltonian for the bound 
state is calculated as 
\begin{equation}
H_k = k \left(\hat{a} \hat{a}^{\dagger} + \frac{1}{2} \right),
\end{equation}
provided we normalise the modes as in (\ref{wrapp}).
Thus, with our normalisation we obtain the free harmonic oscillator picture 
as we should.

\end{document}